\pdfoutput=1
\documentclass[a4paper,11pt]{article}

\usepackage{amsmath}
\usepackage{amssymb}
\usepackage{graphicx}
\graphicspath{{figures/}} %Setting the graphicspath
\usepackage{xfrac}

\usepackage{braket}

\usepackage[utf8]{inputenc}

\usepackage[square,numbers,comma,sort&compress]{natbib}
\usepackage[colorlinks=true,citecolor=blue,linkcolor=purple,urlcolor=purple]{hyperref}
\usepackage{geometry}
\usepackage[capitalise]{cleveref}
\usepackage[usenames,dvipsnames,svgnames,table]{xcolor}
\usepackage{booktabs}
\usepackage{slashed}

\def\varabstract{ }
\def\varkeywords{ }
\def\vararxivnumber{ }
\def\vartitle{ }
\def\varsubtitle{ }
\renewcommand{\title}[1]{\gdef\vartitle{#1}}

\renewcommand{\abstract}[1]{\gdef\varabstract{#1}}
\newcommand{\keywords}[1]{\gdef\varkeywords{#1}}

\newtoks\authtoks
\renewcommand{\author}[2][]{%
	\authtoks=\expandafter{\the\authtoks#2$^{#1}$\ }%
}
\newtoks\affiltoks
\newcommand{\affiliation}[2][]{%
    \affiltoks=\expandafter{\the\affiltoks{\item[$^{#1}$]#2}}%
}
\newtoks\emailtoks\newcounter{emailcounter}%
\newcommand{\emailAdd}[1]{%
\ifnum\theemailcounter>0\emailtoks=\expandafter{\the\emailtoks, \typeemail{#1}}%
\else\emailtoks=\expandafter{\typeemail{#1}}%
\fi
\stepcounter{emailcounter}}
\newcommand{\typeemail}[1]{\href{mailto:#1}{\tt #1}}

\renewcommand\maketitle{
	\newgeometry{margin=2cm}
	\pagestyle{empty}\setcounter{page}{0}
	{\huge\flushleft\sffamily\bfseries\vartitle\\\Large\varsubtitle\par}
\vskip6ex
{\large\bfseries\raggedright\sffamily\the\authtoks\par}
\vskip2ex
\begin{list}{}{%
\setlength{\leftmargin}{0.28cm}%
\setlength{\labelsep}{0pt}%
\setlength{\itemsep}{-3pt}%
\setlength{\topsep}{-\parskip}}
\itshape\small%
\the\affiltoks
\end{list}
\vskip2ex
\noindent\hspace{0.28cm}\begin{minipage}[l]{.9\textwidth}
\begin{flushleft}
\textit{E-mail:} \the\emailtoks
\end{flushleft}
\end{minipage}
\vskip5ex
\noindent{\renewcommand\baselinestretch{.9}\textsc{Abstract:}}\ \varabstract
\vskip5ex
\if!\varkeywords!\else\noindent{\textsc{Keywords:}}\ \varkeywords \vskip2ex\fi
\if!\vararxivnumber!\else\noindent{\textsc{ArXiv ePrint:}} \href{http://arxiv.org/abs/\vararxivnumber}{\vararxivnumber}\vskip2ex\fi

\newpage
\restoregeometry
\pagestyle{plain}
\setcounter{footnote}{0}
}

\usepackage[square,numbers,comma,sort&compress]{natbib}

\definecolor{MS}{rgb}{0,0,1}

\usepackage{ifthen}

	\newcommand{\barlimc}[7]{
  \pgfmathparse{\mypos+0.3}
  \edef\mypos{\pgfmathresult}
		\node[left,scale=0.6] at (0,\mypos) {#1};
		\pgfmathparse{#3 > 5 ? 1 : 0}
		\ifthenelse{\pgfmathresult=1}{
			\fill[#2] ($(0,\mypos)+(0,-0.1)$) rectangle +(5,0.2);
			\fill[white] ($(0,\mypos)+(3.5,-0.1)$) rectangle +(0.3,0.2);
			\draw[decoration={zigzag},decorate,#2,very thick] (3.4,\mypos) to +(0.5,0);
			\node[left,scale=0.6] at (5,\mypos) {#3};
			}{
			\fill[#2] ($(0,\mypos)+(0,-0.1)$) rectangle +(#3,0.2);
			\node[left,scale=0.6] at (#3,\mypos) {#3};
		}
		\fill[#4] ($(0,\mypos)+(0,-0.1)$) rectangle +(#5,0.2);
		\node[left,scale=0.6] at (#5,\mypos) {#5};
		\fill[#6] ($(0,\mypos)+(0,-0.1)$) rectangle +(#7,0.2);
		\pgfmathparse{#7 <0.3 ? 1 : 0}
		\ifthenelse{\pgfmathresult=1}{
			\node[right,scale=0.6] at (0,\mypos) {#7};
		}{
		\node[left,scale=0.6] at (#7,\mypos) {#7};
	}
}

\newcommand{\prd}[3]{{\it Phys. Rev.} {\bf D#1} (19#3) #2}
\newcommand{\pl}[3]{{\it Phys. Lett.} {\bf #1B}  (19#3) #2}
\newcommand{\np}[3]{{\it Nucl. Phys.} {\bf B#1}  (19#3) #2}
\newcommand{\prl}[3]{{\it Phys. Rev. Lett.} {\bf #1}  (19#3) #2}
\newcommand{\tom}{\leftrightarrow}
\newcommand{\ra}{\rightarrow}
\newcommand{\half}{{1\over 2}}
\newcommand{\matel}[3]{\left<#1\right|#2\left|#3\right>}
\newcommand{\cp}{{\cal CP}}
\newcommand{\CP}{{\cal CP}}
\newcommand{\pp}{{\cal P}}
\newcommand{\C}{{\cal C}}
\newcommand{\cpv}{\not\!\!{\cal CP}}
\newcommand{\T}{{\cal T}}
\newcommand{\Tn}{\not\!{\cal T}}
\newcommand{\dfs}{(\phi^I_s-\phi^J_s)}
\newcommand{\dfp}{(\phi^I_p-\phi^J_p)}
\newcommand{\dds}{(\delta^I_s-\delta^J_s)}
\newcommand{\ddp}{(\delta^I_p-\delta^J_p)}
\newcommand{\dfsp}{(\phi^I_s-\phi^J_p)}
\newcommand{\ddsp}{(\delta^I_s-\delta^J_p)}

\title{CONSTRUCTING $\cp$-ODD OBSERVABLES}
\author[1]{German Valencia}%\emailAdd{German.Valencia@monash.edu}
\affiliation[1]{Physics Department, Iowa State University, Ames, Iowa 50011}

\abstract{
In these lectures we review the general features necessary to
construct $\cp$ odd observables and study illustrative examples.
We present in some detail the case
of $\cp$ violation in hyperon decays. We survey different observables
sensitive to $\cp$ violating new physics, concentrating on the search for
electric-dipole moments in high energy experiments.
}

\keywords{CP-odd, T-odd observables}

\begin{document}

\maketitle

{\hypersetup{linkcolor=black}
  \tableofcontents}

\newpage

\section{Discrete Symmetries}

We start by recalling some basic properties of the discrete symmetries
corresponding to parity, charge conjugation and time reversal invariance. The
first two are implemented by unitary transformations, and the last one by
an anti-unitary transformation in free field theories.\cite{tdlee,itzu,gross}

Parity $(\pp)$ is the discrete symmetry that
takes $\vec{x} \tom -\vec{x}$. For free,
single particle, momentum eigenstates, this simply reverses the
particle momentum up to
a phase $\eta = \pm 1$:
\begin{equation}
\pp\ket{\Psi(\vec{k},\vec{s})}=\eta^\pp \ket{\Psi(-\vec{k},\vec{s})} .
\label{par}
\end{equation}
For scalars the intrinsic phase is the same for particle and anti-particle,
whereas for fermions it is opposite. Photons have intrinsic parity $\eta= -1$.
For angular momentum eigenstates, a phase of $(-)^l$ is introduced
by the parity transformation.

Charge conjugation $(\C)$ is the discrete symmetry that
transforms particles into
antiparticles up to a phase. For both scalars and fermions the intrinsic phase
associated with an anti-particle
is the complex conjugate of the intrinsic phase associated with the
corresponding particle. The intrinsic phase for photons is $\eta = -1$. The
phase associated with each particle is convention dependent and
unphysical.

The discrete symmetry corresponding to a successive application of $(\pp)$
and $(\C)$, $(\cp)$, is the subject of these lectures. Some examples of a
$\cp$ transformation on common single particle states within our phase
convention are:
\begin{eqnarray}
\CP \ket{\pi^0}& = & - \ket{\pi^0} \nonumber \\
\CP \ket{\pi^\pm}& = & - \ket{\pi^\mp} \nonumber \\
\CP \ket{K^0}& = & - \ket{\bar{K}^0} .
\label{cpmeson}
\end{eqnarray}
An important system is that of
a fermion anti-fermion pair in their center of mass frame. This system
transforms under $\cp$ as:
\begin{eqnarray}
\CP \ket{e^-(\vec{k},\vec{s})e^+(-\vec{k},\vec{s}^\prime )} & = &
\C(-)\ket{e^-(-\vec{k},\vec{s})e^+(\vec{k},\vec{s}^\prime )} \nonumber \\
 & = & (-)\ket{e^+(-\vec{k},\vec{s})e^-(\vec{k},\vec{s}^\prime )} \nonumber \\
 & = & \ket{e^-(\vec{k},\vec{s}^\prime)e^+(-\vec{k},\vec{s})}
\label{cpfpair}
\end{eqnarray}
the net effect of the  $\CP$ transformation is to interchange the spin vectors
of particle and anti-particle. This result implies that it is possible to
construct simple $\cp$-odd observables for processes that occur in $e^+ e^-$ or
$p\overline{p}$ colliders when the spin density matrix for the initial
state is symmetric under a spin interchange.\cite{donva,dhv}
We will use this transformation repeatedly
in the construction of $\CP$ odd observables.

Time reversal invariance $({\cal T})$ is the symmetry that takes $t \tom -t$
classically. The requirement that under this transformation a Hamiltonian
does not change sign, leads to the anti-unitary nature
of ${\cal T}$.\cite{gross} For free, single particle, momentum
eigenstates, the effect of this transformation is to
reverse the direction of both momentum and spin vectors.
There is also an intrinsic
phase associated with this transformation. It is the same for particle and
anti-particle. The anti-unitary nature of the transformation is responsible
for the additional effect of interchanging incoming states into
outgoing states and viceversa:
\begin{equation}
{\cal T} \ket{p(\vec{k},\vec{s})} = \bra{p(-\vec{k},-\vec{s})}
\equiv \bra{\tilde{p}}.
\label{time}
\end{equation}
We will use the symbol $\tilde{p}$ to denote
a reversal of all momenta and spin vectors in the state $p$.

The anti-unitary nature of the time reversal operator is also important when
applying this transformation to multi-particle states.
If the multi-particle state
is a direct product of free (non-interacting, asymptotic) momentum eigenstates,
then
the $\T$ transformation is a straightforward generalization of Eq.~\ref{time}.
For angular momentum eigenstates, the transformation is:
\begin{equation}
{\cal T}\ket{j,m} = (-)^{j+m} \ket{j,-m}.
\label{timeang}
\end{equation}
However, if the multi-particle state consists of interacting particles,
the interchange of ``in'' and ``out'' states plays a crucial role, as we will
see in the example of hyperon decay. In general, the transformation is:
\begin{equation}
{\cal T}\ket{f_{out}} = \bra{\tilde{f}_{in}}=\bra{\tilde{f}_{out}}S^\dagger .
\label{inout}
\end{equation}
The action of the anti-unitary operator ${\cal T}$ can be expressed in terms
of that of a unitary operator $U_T$ following Wigner:\cite{tdlee,wigner}
\begin{eqnarray}
{\cal T}\ket{n} &=& U_T \ket{n}^* \nonumber \\
{\cal T} {\cal O} {\cal T}^{-1} &=& U_T {\cal O}^* U_T^\dagger .
\label{wigner}
\end{eqnarray}

It is convenient to define an operator $(T)$ sometimes called
``naive''-time reversal transformation. This operator which is
{\it not} the same as the time reversal operator ${\cal T}$, simply
reverses the sign of all momentum and spin vectors:
$T\ket{p}=\ket{\tilde{p}}$. $T$  is a useful
operator to classify $\CP$ odd observables.

Throughout these lectures we will assume that the combined operation
of parity, charge conjugation and time reversal, $\CP{\cal T}$, is a
good symmetry
of the theories under consideration. As is well known from the $\CP{\cal T}$
theorem, this will be true for any theory defined by a hermitian, Lorentz
invariant, normal ordered product of fields quantized with the usual
spin-statistics connection.\cite{itzu}
Because of this assumption, $\CP$ violation
(denoted by $\cpv$) will be equivalent to ${\cal T}$ violation
($\Tn$).

\section{Ingredients for $\cpv$ Observables}

A $\CP$-odd observable is an observable whose expectation value vanishes
if $\cp$ is conserved. There are several ingredients necessary to
construct $\cp$-odd observables.

The first of these ingredients is to have a $\CP$ violating phase in the
theory. The theory must have a non-trivial phase (one that cannot be removed
by field redefinitions) in order to violate $\CP$. As you already saw in
lectures by previous speakers, the minimal standard model with three
generations has one such phase.\cite{rafael}
This phase appears in the CKM matrix, and in the
original parameterization of Kobayashi and Maskawa it
is called $\delta$.\cite{kobma}

Once we have a theory that contains a $\CP$ violating phase, we must
still construct an observable that depends on the value of that phase.
One  possibility is to identify a process that is forbidden by $\CP$
invariance, typically a transition between $\CP$ eigenstates with
different $\CP$ eigenvalues. Most processes, however, involve states
that are not $\cp$ eigenstates. In this case $\cp$ invariance predicts
relations between the process and its $\CP$ conjugate process. One can
then construct observables that test these predictions.

A $\cpv$ observable that compares a pair of $\cp$ conjugate processes,
will vanish unless there are several amplitudes contributing to the
processes, and these amplitudes can interfere. This can be seen
by considering the process $i \ra f$ and assuming that
there is only one amplitude
contributing to it, so that $M(i\ra f) = e^{i\phi}A_1$. Even if this amplitude
contains the $\CP$ violating phase $\phi$, observables will be proportional to
$|M|^2=A_1^2$ and thus independent of $\phi$. However, if there is at least one
other amplitude (with a different phase), then $M(i\ra f) =e^{i\phi}A_1 +A_2$,
and observables will be proportional to $|M|^2=(A_1\cos\phi +
A_2)^2+A_1^2\sin^2
\phi$. Interference is, therefore, necessary for observables to depend on
$\cp$ violating phases. However, we do not yet have $\cp$-odd observables,
since $|M|^2$ above does not vanish as $\phi \ra 0$. It is still possible to
extract information on $\CP$ violation from precision measurements
of these observables. A known example is the
determination of the CKM parameters by measurements of the {\it sides}
of the unitarity triangle.\cite{buras}

The construction of $\CP$-odd observables (that vanish when $\phi\ra 0$)
requires an additional (usually $\CP$-conserving) phase. A very simple way
to see this, is to rewrite the previous amplitude for our process $i \ra f$
as $M(i\ra f)=A_1\cos\phi+A_2+iA_1\sin\phi$. One can then see immediately,
that it is only possible to obtain a term linear in $\sin\phi$ in the
matrix element squared $|M|^2$, if there is another imaginary term in $M$
to interfere with the $iA_1\sin\phi$ term. If, for example, there is
another $\CP$ conserving amplitude $A_3 e^{i\delta}$ contributing to
our process $i \ra f$, it will be possible to find observables proportional
to $\sin\phi\sin\delta$. A $\CP$ conserving phase such as this, usually
arises from final state interactions (from the existence of real intermediate
states beyond the Born approximation). You have already seen an example of
this type of phase in the study of $\CP$ violation in the $K^0 \ra \pi \pi$
system\cite{rafael}. In that case you wrote, for
example, $A(K^0 \ra \pi^+ \pi^-)=A_0e^{i\delta_0}+A_2/\sqrt{2}
e^{i\delta_2}$. In this expression, any $\CP$ violating phases are included
in $A_0,A_2$, and the phases $\delta_0$, $\delta_2$ are the $\pi \pi$
scattering phase shifts in the $I=0$, $I=2$ channels.

Using the discrete symmetry $T$ (recall that this is {\it not} the same as
time reversal invariance) to classify $\CP$-odd observables, one finds in
general that $\CP$-odd and $T$-even observables are proportional to
quantities like $\sin\delta\sin\phi$. That is, that the simultaneous presence
of a $\CP$ violating phase and a $\CP$ conserving ``unitarity'' phase is needed
in order for the observable not to vanish. On the other hand,
$T$-odd observables are found to be proportional to quantities like
$\sin\delta\cos\phi+\sin\phi\cos\delta$. This result is consistent with the
statement that $T$ violation is {\it not} the same as $\CP$ violation. Using
$T$-odd quantities it is possible to construct $\CP$-odd observables
that do not require additional ``unitarity'' phases.

We will see how this works in more detail when we discuss specific examples
later on.
At this point, however, it is convenient to ask two questions.
\begin{itemize}
\item In view
of our earlier discussion, how is it possible to get $T$-odd and $\CP$-odd
observables that do not require a ``unitarity'' phase? A typical $T$-odd
quantity contains the triple product of three vectors (recall that $T$
reverses the sign of momentum and spin vectors), like $\vec{p_1}\cdot
\vec{p_2}\times\vec{p_3}$. Such a quantity must come from a Lorentz invariant
expression of the form $\epsilon^{\mu\nu\alpha\beta}k_\mu p_{1\nu} p_{2\alpha}
p_{3\beta}$ (for example in the rest frame of $k$). In the case of amplitudes
involving fermions, we would obtain an expression like this one from the
Dirac trace of four $\gamma$ matrices and a $\gamma_5$, which has a factor of
$i$ relative to expressions without the epsilon tensor. This relative phase
is the one taking the place of the ``unitarity'' phase
needed for a $\CP$-odd observable.

\item Why is it possible to
obtain a $T$-odd observable that does not violate $\CP$? As we said before,
this is simply because $T$ is {\it not} the same as time reversal invariance.
In some detail, we can see the origin of such $T$-odd observables when we
go beyond the Born approximation for a given process. Writing the $S$-matrix
as $S=I-iM$, unitarity implies that
\begin{eqnarray}
S^\dagger S = I &=& (I+iM^\dagger)(I-iM) \nonumber \\
&=& I-i(M-M^\dagger )+M^\dagger M
\nonumber \\ \Rightarrow \;\;
i(M-M^\dagger ) &=& M^\dagger M.
\label{usma}
\end{eqnarray}
Taking the matrix element of this last expression between states $f$ and $i$,
with the definition $M_{if} \equiv \bra{f}M\ket{i}$, and noticing that
$\bra{f}M^\dagger\ket{i}=\bra{i}M\ket{f}^*$ one finds:
\begin{equation}
i(M_{if}-M^*_{fi})=\sum_nM^*_{fn}M_{in}.
\label{intstate}
\end{equation}
This shows that for the left hand side to be different from zero,
there must exist real intermediate states, $n$, that couple to
$i$ and $f$. This is shown schematically
in Figure~\ref{unit}.
\begin{figure}[htp]
\centering{\includegraphics[width=12cm]{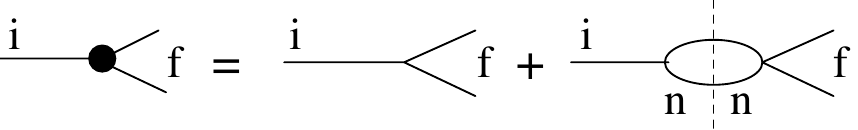}}
\caption{Unitarity phases from real intermediate states.}
\label{unit}
\end{figure}

Assuming time reversal invariance (and hence $\CP$ invariance) it is possible
to relate the matrix elements for the process $i \ra f$ and for its $T$
conjugated process $\tilde{i} \ra \tilde{f}$:
\begin{equation}
\bra{f}M\ket{i}=\bra{f}{\cal T}^{-1}{\cal T} M {\cal T}^{-1} {\cal T}\ket{i} =
\eta \bra{\tilde{i}}M\ket{\tilde{f}}.
\label{uni}
\end{equation}
We have allowed for a possible phase $\eta$ associated with the
time reversal operation. If we write $M$ as:
\begin{equation}
M= {1 \over 2}(M + M^\dagger)+{1\over 2}(M-M^\dagger),
\label{hdec}
\end{equation}
we can rewrite Eqn.~\ref{uni} as:
\begin{equation}
\bra{f}(M+M^\dagger)\ket{i}+\bra{f}(M-M^\dagger)\ket{i}=
\eta\biggl( \bra{\tilde{f}}(M+M^\dagger)\ket{\tilde{i}}^*-
\bra{\tilde{f}}(M-M^\dagger)\ket{\tilde{i}}^*\biggr).
\label{tborn}
\end{equation}
{}From this expression, it is clear that in the Born approximation,
where $\bra{f}(M-M^\dagger)\ket{i}=0$, time reversal invariance
implies $T$ conservation (thus the name ``naive'' time reversal
invariance sometimes used for $T$). It is also clear, that beyond the
Born approximation, $\bra{f}(M-M^\dagger)\ket{i} \not= 0$,
it is possible to construct $T$-odd observables even when time
reversal invariance is a good symmetry.\cite{gavela}

\end{itemize}

\section{How large can $\cpv$ be?}

In this section we discuss how large the $\cpv$ signals
can be. We recall the unitarity bounds on $\cp$ violating
phases in a few models. We also discuss generalities of other
suppression factors and illustrate this with an example.

\subsection{Minimal Standard Model}

In the minimal standard model at least three generations of quarks are
required to accommodate $\cp$ violation.\cite{kobma}
Even if there are three generations,
there will only be $\cp$ violation if no two quarks of the same charge are
degenerate in mass, and if all three generations mix; that is, no angle in
the CKM matrix can be $0$ or $\pi/2$. A convenient way to express this, is to
note that all $\cp$ violation is proportional to\cite{jarskog,ddwu}:
\begin{equation}
J={\rm Im}\left(V_{us}V_{cb}V^*_{ub}V^*_{cs}\right)
={\rm Im}\left(V_{ud}V_{tb}V^*_{ub}V^*_{td}\right)
\label{jinv}
\end{equation}
In the Wolfenstein parameterization of the  CKM matrix $J=A^2\lambda^6\eta$.
With present constraints, $\lambda =0.22$, $A=0.9 \pm 0.1$,
$\sqrt{\rho^2+\eta^2}=0.4\pm 0.2$ one finds:
\begin{equation}
|J| \leq 6.8 \times 10^{-5}.
\label{jexp}
\end{equation}
It is instructive to compute $J$ in the original KM parameterization:
\begin{equation}
J=\cos\theta_1\cos\theta_2\cos\theta_3\sin^2\theta_1\sin\theta_2
\sin\theta_3\sin\delta.
\label{jkm}
\end{equation}
{}From this expression, we can find the maximum value that $J$ can take.
This parameterization in terms of cosines and sines of angles follows
from three generation unitarity of the CKM matrix. Therefore, the maximum
value that the expression for $J$ can take:
\begin{equation}
J_{max} ={1 \over 6 \sqrt{3}} \approx 0.1
\label{jkmmax}
\end{equation}
is referred to as the ``unitarity'' upper bound on $\cp$ violation.
The purpose of this exercise is to see that in the CKM model of
$\cp$ violation, the experimentally allowed upper bound for
$\cp$ violation is at least three orders of magnitude smaller than the
theoretical ``unitarity'' upper bound. This must be kept in mind
when using unitarity upper bounds on $\cp$ violating parameters in
other models of $\cp$ violation to estimate the potential size of
observables.

For $\cp$ violation there is also the condition of non-degeneracy
of quark masses. It is tempting to write this condition by saying that
$\cp$ violation must be proportional to the factor:
\begin{equation}
{(m_t^2-m_c^2)(m_t^2-m_u^2)(m_c^2-m_u^2)
(m_b^2-m_s^2)(m_b^2-m_d^2)(m_s^2-m_d^2)\over M^{12} }
\label{massfac}
\end{equation}
where $M$ is some typical mass in the problem. This, however, is
{\it not} true. In general, the requirement that quarks of the same
charge cannot be degenerate is fulfilled in a more subtle way.

\subsection{Suppression factors in $Z \ra d_i \overline{d}_j$}

\begin{figure}[htp]
\centering{\includegraphics[width=10cm]{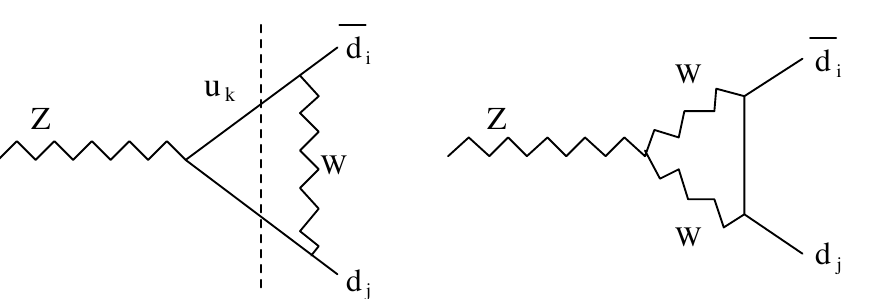}}
\caption[]{Some diagrams contributing to the asymmetry defined in
Eq.~\ref{deltaz}.}
\label{zas}
\end{figure}

Let us consider the case of flavor changing decays of the $Z$ boson as
an example. It is possible to construct a $\cp$ odd rate asymmetry:
\begin{equation}
\Delta \equiv {\Gamma(Z \ra \overline{b}s)-\Gamma(Z \ra \overline{s}b)
\over \Gamma(Z \ra \overline{b}s) + \Gamma(Z \ra \overline{s}b)}
\label{deltaz}
\end{equation}
Clearly, this is a $T$-even observable, and
according to our previous discussion it will vanish unless there are
non-zero unitarity phases.
In the standard model, in unitary gauge, the amplitude for the
process $Z \ra d_i \overline{d}_j$ would be computed
from the diagrams in Figure~\ref{zas} (plus all other related ones).
It can be seen from Figure~\ref{zas}, that there can be real
intermediate states
when the intermediate $U_k$ quark is either an up or charm-quark.
In this case, the
first diagram has an absorptive part that provides the unitarity phase.
The rate can be computed as a sum of contributions from
each of the three possible intermediate quarks in the first diagram.
The authors of Ref.\cite{bernabeu} find:
\begin{equation}
\Gamma(Z \ra d_i \overline{d}_j)={g^2\over \cos^2\theta_W}{M_Z \over 8 \pi}
\left|{g^2 \over 2^5 \pi^2}\right|^2\left|\sum_k \xi_k
I(r_k,s)\right|^2,
\label{gavres}
\end{equation}
where $r_k =m_k^2/M_W^2$, $s=M_Z^2/M_W^2$ and $\xi_k = V^*_{ki}V_{kj}$,
and the $I(r_k,s)$ are the form factors obtained after the one-loop
calculation.
Using the unitarity of the CKM matrix, $\xi_1 +\xi_2 +\xi_3 =0$, and defining
$F_2=I_2-I_1$, $F_3=I_3-I_1$, the rate asymmetry is found to be\cite{bernabeu}:
\begin{equation}
\Delta = {-4 {\rm Im}\left(\xi_2\xi_3^*\right){\rm Im}\left(F_2 F^*_3\right)
\over \left|\xi_2F_2+\xi_3 F_3\right|^2+\left|\xi^*_2F_2+\xi_3^*F_3\right|^2}.
\label{gavrescp}
\end{equation}
As expected, the rate asymmetry is proportional to the $\cpv$ phase in the
CKM matrix ${\rm Im}\left(\xi_2\xi_3^*\right)\sim J$. It is also proportional
to the absorptive phase in the loop form factors
${\rm Im}\left(F_2 F^*_3\right)$. This absorptive phase has a simple form
in the limit $r_u <<1$, $r_c << 1$:\cite{bernabeu}
\begin{equation}
{\rm Im}F_2 F_3^* \approx{m_c^2 -m_u^2 \over M_W^2}.
\label{gavsimp}
\end{equation}
A numerical calculation of $\Delta$, with typical values of the CKM angles
and quark masses, yields:\cite{bernabeu}
\begin{eqnarray}
{\Gamma(Z \ra \overline{b}s) \over \Gamma(Z \ra \overline{s}s)} &\approx&
10^{-7} \nonumber \\
\Delta &\approx & 10^{-5}
\label{numgav}
\end{eqnarray}
To understand how this result obtains, we can use the fact that the rate
(and thus the denominator of Eq.~\ref{gavrescp}) is dominated by the
top-quark intermediate state. The CKM factors that enter into the rate
are thus,
\begin{equation}
\Gamma(Z \ra \overline{b}s) \sim |V_{ts}V^*_{tb}|^2 \sim A^2 \lambda^4
\label{ckmfing}
\end{equation}
Assuming that $F_3$ is of order one (it will depend on $m_t^2/M_W^2$), the
size of the rate asymmetry can be understood as:
\begin{equation}
\Delta \sim {(m_c^2-m_u^2) \over M_W^2}{A^2 \lambda^6 \eta \over A^2 \lambda^4}
\sim {m_c^2 \over M_W^2} \lambda^2 \eta \sim 10^{-5}
\label{delcount}
\end{equation}
The lessons to be learned from this example are:

\begin{itemize}

\item Part of the smallness of $J$ that comes into every $\cp$ odd
observable in the standard model, can be compensated by looking
at rare decays (in this case a full factor of $A^2 \lambda^4$). The
price to pay if one wants to look for large asymmetries, is that
one has to look at very rare processes.

\item The requirement that quarks of the same charge not be degenerate,
acts as an additional suppression factor for high energy processes. In this
case, after using the approximation $m_u=0$, we were left with a ratio of the
charm-quark mass to the $Z$-mass.

\item Not all the $m_i \not =m_j$ conditions appear as explicit mass ratios
in the expression for an observable. In this case we do not see any mass
factors associated with down-type quarks. This is because the down-type quarks
in this problem appear as the external states. The condition that they cannot
be degenerate has been used in the definition of the final states.

\end{itemize}

\subsection{Models with extra scalars: basic features}

We will not have time to discuss in detail any model with additional scalars.
Some discussion of the model with two scalar doublets can be found in the
lectures by S.~Dawson. The important feature of the two doublet model for us,
is that one can introduce additional $\cp$ violation in the scalar sector only
at the expense of introducing tree-level flavor changing neutral
currents.\cite{cherev} To get around
this problem one needs a model with at least three scalar doublets, or one
with two doublets and additional singlets. These models turn out to have
too many free parameters to be predictive. As an example let us look at the
Weinberg three doublet model.\cite{weinor} The scalars are:
\begin{equation}
\phi_i = \left(\begin{array}{c}
               \phi_i^+ \\
               \phi_i^0 \end{array}\right),\; i = 1,2,3
\label{scalars}
\end{equation}
There are 12 scalar fields. Three of these 12 become the longitudinal
components of the $W^\pm$ and $Z$ gauge bosons, and we are left with 9 physical
scalars: $H_{1,2}^\pm$ and 5 neutral particles. Thus, there are many possible
additional phases. If we concentrate for the time being in the charged sector,
we can write the transformation between the original scalars and the mass
eigenstate basis as:
\begin{equation}
\left( \begin{array}{c} \phi_1^+ \\ \phi_2^+ \\ \phi_3^+ \end{array}\right)
= Y \left(\begin{array}{c} \omega^+ \\ H_1^+ \\H_2^+ \end{array}\right)
\label{physca}
\end{equation}
where $Y$ is a three by three unitary matrix analogous to the CKM matrix.
This matrix can be parameterized by three angles and one phase \cite{alstye}
in a similar fashion to the CKM matrix. The couplings to fermions are given
by\cite{alstye}:
\begin{equation}
{\cal L}^+_Y=\left( 2\sqrt{2}G_F\right)^{1\over 2}\sum_{i=1}^2\left(
\alpha_i\overline{U}_L V m_D D_R + \beta_i \overline{U}_R m_U V D_L
+\gamma_i \overline{N}_L M_E E_R \right) H_i^+ + {\rm h.~c.}
\label{yukawa}
\end{equation}
In this expression, $V$ is the CKM matrix, $m_D,m_U,m_E$ are mass matrices
for the down-type quarks, up-type quarks and charged leptons. There are
three vacuum expectation values, given in terms of
the usual $v \approx 246$~GeV by:
\begin{eqnarray}
v_1 &=& \tilde{c}_1 v \nonumber \\
v_2 &=& \tilde{s}_1\tilde{c}_2 v \nonumber \\
v_3 &=& \tilde{s}_1\tilde{s}_2 v
\label{vevs}
\end{eqnarray}
where the notation $\tilde{s}_1$ etc. stands for sines and cosines of the
mixing angles in the $Y$ matrix, in complete analogy with the CKM angles.
Imposing a discrete symmetry that removes tree-level flavor changing neutral
currents the couplings in Eq.~\ref{yukawa} are:\cite{alstye}
\begin{eqnarray}
&& \alpha_1 = -\tilde{s}_1 \tilde{c}_3 {v \over v_1} \;\;
\alpha_2 = -\tilde{s}_1 \tilde{s}_3 {v \over v_1} \nonumber \\
&& \beta_1 = [-\tilde{c}_1 \tilde{c}_2\tilde{c}_3
+ \tilde{s}_2\tilde{s}_3e^{i\tilde{\delta}}] {v \over v_2} \;\;
 \beta_2 = [-\tilde{c}_1 \tilde{c}_2\tilde{s}_3
- \tilde{s}_2\tilde{c}_3e^{i\tilde{\delta}}] {v \over v_2} \nonumber \\
&& \gamma_1 = [\tilde{c}_1 \tilde{s}_2\tilde{c}_3
+ \tilde{c}_2\tilde{s}_3e^{i\tilde{\delta}}] {v \over v_3} \;\;
 \gamma_2 = [\tilde{c}_1 \tilde{s}_2\tilde{s}_3
- \tilde{c}_2\tilde{c}_3e^{i\tilde{\delta}}] {v \over v_3}
\end{eqnarray}
By analogy with the CKM matrix, one can easily show that all $\cp$ violation
will be proportional to:\cite{cherev}
\begin{equation}
J_W=\cos\tilde{\theta}_1\cos\tilde{\theta}_2
\cos\tilde{\theta}_3\sin^2\tilde{\theta}_1\sin\tilde{\theta}_2
\sin\tilde{\theta}_3\sin\tilde{\delta}
= \left({v_1 v_2 \over v^2}\right)^2 {\rm Im}\alpha_1\beta^*_1
\label{jhiggs}
\end{equation}
and there is, of course, a unitarity upper bound:
\begin{equation}
J_W \leq {1 \over 6 \sqrt{3}}\approx 0.1.
\label{jhuni}
\end{equation}
Many estimates of $\cp$ odd observables in this model have been made
using the unitarity upper bound for the unknown mixing angles. In
this regard, it is convenient to keep in mind the situation in the
CKM model of $\cp$ violation, where the maximum experimentally allowed
value of $J$ is three orders of magnitude smaller than the unitarity
upper bound. In the case of the Weinberg three doublet model, we can
see that the unitarity bound is achieved if $\tilde{c}_1 = 1/\sqrt{3}$
and $\tilde{c}_2 = 1/\sqrt{2}$, so that $v_1/v_2 = \sqrt{2/3}$.
However, the model was constructed so that the up-type quarks get their mass
from $v_2$ and the down-type quarks get their mass from $v_1$. One might
therefore expect that $v_1/v_2 \sim m_b/m_t << 1$. Of course, this need
not be the case, but it just emphasizes that there is no reason for the
mixing angles to be such as to maximize the $\cp$-odd invariant $J_W$.

As we said before, this model also contains five physical neutral scalars.
If there is $\cp$ violation in the charged sector, it is natural
to find it in the neutral sector as well.\cite{deshpande}
One could proceed in a manner
analogous to what we did for the charged sector, but things would be more
complicated by the larger number of particles involved. It is conventional
to assume that low energy observables will be dominated by the effects of the
lightest neutral field, called $H^0$. This field couples to fermions in a way
given by:
\begin{equation}
{\cal L}=-\left({m_f \over v}\right) H^0 \overline{f}\left(
AP_L + A^* P_R\right) f = -\left({m_f \over v}\right) H^0
\overline{f}\left( {\rm Re}A-i {\rm Im}A\gamma_5 \right)f.
\label{neuyuk}
\end{equation}
This form exhibits the general result that the simultaneous
presence of scalar and pseudo-scalar couplings of a neutral
spinless field to fermions, signals the violation of $\cp$.
Weinberg has performed a general analysis of unitarity bounds,\cite{wein}
and assuming that $v_1 \sim v_2 \sim v_3$ the limit in this case
reads:
\begin{equation}
\left|{\rm Im}A^2\right| < \sqrt{2}.
\label{unithn}
\end{equation}
This limit has been used extensively in the literature to estimate
the size of potential $\cp$ violating observables, and we shall use
it in the examples that will be discussed later. Once more, however,
we should keep in mind that there is no reason why the mixing angles
would be such as to yield the largest possible $\cp$ violation.

In general, one finds that models of $\cp$ violation beyond the
minimal standard model contain large numbers of parameters and
new particles. It is reasonable to think that the first experimental
evidence for this kind of model would be the actual discovery of one
of the new particles. To search for evidence for this type of new
physics through $\cp$ violation, only makes sense if the experiments
are being performed at energies much lower than the threshold for
production of the new particles. In this case we can treat the new
particles as heavy, and discuss their low energy effects in terms
of an effective $\cp$ violating (or conserving) Lagrangian that
contains only the fields of the minimal standard model.\cite{burg}
This would be
a non-renormalizable Lagrangian, with arbitrary coupling constants
that parameterize the effects of the heavy physics. It is in this
context that we will discuss several examples later on.

We now turn our attention to the construction of examples of $\CP$-odd
observables of different types.

\newpage
\section{$K_L \ra \pi^0 \nu \overline{\nu}$}

In this section we will study one of the most interesting rare kaon
decays, $K_L \ra \pi^0 \nu \overline{\nu}$.\cite{lit}
{}From the perspective of
our previous discussion, this is an example of a process that is
forbidden by $\CP$ (at least to a very good approximation), and hence
its simple observation would signal the violation of $\CP$.

\begin{figure}[htp]
\centering{\includegraphics[width=12cm]{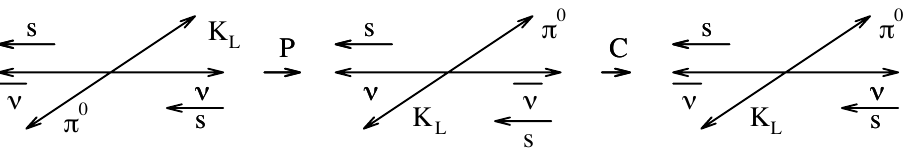}}
\caption{$K_L \ra \pi^0 \nu \overline{\nu}$}
\label{kpnn}
\end{figure}

\begin{figure}[htp]
\centering{\includegraphics[width=12cm]{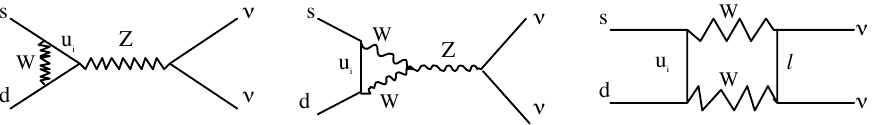}}
\caption{$K_L \ra \pi^0 \nu \overline{\nu}$}
\label{kpnndia}
\end{figure}

We can see how this works by looking at this decay in the
$\nu \overline{\nu}$ center of mass frame. To first order in the
weak interactions, the neutrino pair is in a $J=1$ state, so the
configuration of momenta and spin can look as in Fig.~\ref{kpnn}.
As sketched in that figure, the reaction transforms into itself
under $\CP$. Neglecting $\CP$ violation in the neutral
kaon mass matrix, the initial state has $\CP$ eigenvalue
$\eta^{\CP}_{K_L}=(-)$. The final state has a $\CP$ eigenvalue
$\eta^{\CP}_{\pi^0} (-)(\eta^\pp_\nu)^2(\eta^\C_\nu)^2 = (+)$,
where we have used the properties of the $\pp$ and $\C$ phases
discussed in Section~1.
This means that observation of this decay at a level consistent
with a transition of first order in the weak interactions, would
be an unambiguous indication of $\CP$ violation.
In the minimal standard model this process occurs through the
diagrams of Figure~\ref{kpnndia}. They have been calculated
by several authors, and the salient feature is that the
amplitude is dominated by the top-quark intermediate state.\cite{buras}
One finds that the direct $\CP$ violating amplitude is much
larger than the indirect $\CP$ violating amplitude originating
in the $\CP$ even
component of the initial $K_L$ state. A convenient form to
write the final result is:\cite{buras}
\begin{eqnarray}
B(K_L \ra \pi^0 \nu \overline{\nu}) \approx 1.3 \times 10^{-10}
\eta^2 \biggl[{x_t \over 8}\biggl({x_t+2 \over x_t -1} +
{3x_t-6 \over (x_t-1)^2}\log x_t\biggr)\biggr]^2
\label{bkpnn}
\end{eqnarray}
where $x_t=M_t^2/M_W^2$ and $\eta$ is the imaginary element of
the CKM matrix in the Wolfenstein parameterization.\cite{wolfkm}
You can see
in this formula that the rate is directly proportional to $\eta^2$,
that is, it vanishes if there is no $\CP$ violation.

Experimentally it is very difficult to observe this decay. Taking
the different standard model parameters to lie in their current
allowed ranges gives a branching ratio around $10^{-10}$ to $10^{-11}$.
This is roughly the size of the smallest limit ever placed on a
rare kaon decay $B(K_L \ra \mu^\pm e^\mp)\leq 3.3 \times 10^{-11}$.
However, the decay $K_L \ra \pi^0 \nu \overline{\nu}$ is much more
difficult to study because the two neutrinos are missed. Currently,
the best bound for this mode comes from FNAL-731
$B(K_L \ra \pi^0 \nu \overline{\nu})\leq 2.2 \times 10^{-4}$ although
a proposed experiment, FNAL-799, expects to reach the $10^{-8}$ level
of sensitivity.\cite{liva}
\newpage
\section{Hyperon decays: $\Lambda^0 \ra p \pi^-$}

In this section we will study in detail the reaction $\Lambda^0 \ra p \pi^-$.
This will serve as an example of a sufficiently complicated system that allows
the construction of many $\CP$-odd observables. In Table~\ref{t hyp} we
summarize the quantum numbers and naive quark model content
of the particles we consider.

\begin{figure}[htp]
\centering{\includegraphics[width=10cm]{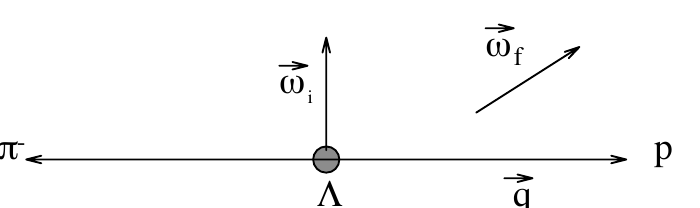}}
\caption{Kinematics of $\Lambda^0 \ra p \pi^-$ in the $\Lambda^0$ rest frame}
\label{lppi}
\end{figure}

\begin{table}[htb]
\centering
\caption[]{Properties of the particles in $\Lambda \ra N \pi$.}
\begin{tabular}{|c|c|c|c|} \hline
Particle & Quark Content &  $J^P$  &  I \\ \hline
$\Lambda^0$ & $uds$ & $\half^+ $ & 0 \\
$p$ & $uud$ & $\half^+$ & $\half$ \\
$n$ & $udd$ & $\half^+$ & $\half$ \\
$\pi^-$ & $\overline{u}d$ & $0^-$ & 1 \\
$\pi^+$ & $\overline{d}u$ & $0^-$ & 1 \\
$\pi^0$ & ${1 \over \sqrt{2}}(\overline{u}u-\overline{d}d)$
& $0^-$ & 1 \\ \hline
\end{tabular}
\label{t hyp}
\end{table}
In Figure~\ref{lppi} we define the notation to be used for the kinematics
of the reaction in
the $\Lambda^0$ rest frame.
In the $\Lambda^0$ rest frame, $\vec{\omega}_{i,f}$ represent unit vectors in
the directions of the $\Lambda$ and $p$ polarizations, and
$\vec{q}$ is the proton momentum.
It can be seen that the final state has isospin
$I= 1/2 \oplus 3/2 = 1/2 {\rm ~or~} 3/2$. Since the initial particle has
isospin $0$, the final state with isospin $1/2$ can be reached via the
$\Delta I = 1/2$ weak Hamiltonian and the final state with isospin $3/2$
can be reached
via the $\Delta I = 3/2$ weak Hamiltonian. The final state has an orbital
parity $(-)^l$ with $l$ being the orbital angular momentum. From conservation
of angular momentum we see that the final state must have $J=1/2$, so that
there are two possible values of $l$. They correspond to the
two possible parity
states: the $s$-wave, $l=0$, parity odd state (thus reached via a parity
violating amplitude); and the $p$-wave, $l=1$, parity even state reached
via a parity conserving amplitude.

By considering the measured branching ratios:
\begin{eqnarray}
B(\Lambda^0 \ra p \pi^-) &=& 64\% \nonumber \\
B(\Lambda^0 \ra n \pi^0) &=& 36\% ,
\label{lbr}
\end{eqnarray}
and the pion-nucleon system in an isospin basis:
\begin{eqnarray}
\ket{I={\half},\; I_3= -{\half}}
&=& \sqrt{1 \over 3}\ket{n\pi^0} - \sqrt{2 \over 3}
\ket{p\pi^-} \nonumber \\
\ket{I={3\over 2},\; I_3= -{\half}}
&=& \sqrt{2 \over 3}\ket{n\pi^0} + \sqrt{1 \over 3}
\ket{p\pi^-},
\label{liso}
\end{eqnarray}
we can see that the final state is mostly an isospin $I=1/2$ state. This
is the empirical result known as the $\Delta I = 1/2$ rule.\cite{marshak}

We first perform a model independent analysis of the decay by writing the
most general matrix element consistent with Lorentz
invariance:\cite{marshak,commins}
\begin{equation}
{\cal M} = G_F m_\pi^2 \overline{U}_P(A-B\gamma_5)U_\Lambda .
\label{gmatel}
\end{equation}
Because this is a two-body decay, the kinematics is fixed and $A$ and $B$
are simply constants. Other forms one could write, can be absorbed into
$A$ and $B$, for example:
\begin{equation}
q_\mu \overline{U}_P \gamma_\mu U_\Lambda = m_P
\overline{U}_P  U_\Lambda
\label{glimat}
\end{equation}
is equivalent to a contribution to $A$. The form Eq.~\ref{gmatel}
can be derived from an effective Lagrangian like:
\begin{equation}
{\cal L} \sim \overline{\Psi}_P ( A - B\gamma_5) \Psi_\Lambda \pi .
\label{lagp}
\end{equation}
Recalling the transformation properties of fermion bilinears and pions
under parity:\cite{itzu,gross}
\begin{eqnarray}
\overline{\Psi}_P \Psi_\Lambda & \stackrel{P}{\longrightarrow} &
\overline{\Psi}_P \Psi_\Lambda  \nonumber \\
\overline{\Psi}_P \gamma_5 \Psi_\Lambda & \stackrel{P}{\longrightarrow} &
-\overline{\Psi}_P \gamma_5 \Psi_\Lambda  \nonumber \\
\pi & \stackrel{P}{\longrightarrow} & -\pi ,
\label{ptrans}
\end{eqnarray}
one finds that
\begin{equation}
{\cal L} \stackrel{P}{\longrightarrow}
\overline{\Psi}_P ( -A - B\gamma_5) \Psi_\Lambda \pi .
\label{ptlag}
\end{equation}
Thus, $A$ corresponds to the parity violating, $s$-wave amplitude; and
$B$ to the parity conserving, $p$-wave amplitude.

To proceed, we write explicit expressions for the
spinors in the $\Lambda$ rest frame:
\begin{equation}
U_\Lambda = \sqrt{2M_\Lambda}\left( \begin{array}{c}
1 \\ 0 \end{array} \right) \chi_i \;\;
U_P = \sqrt{E_P+M_P}\left( \begin{array}{c} 1\\
{\vec{\sigma}\cdot\vec{q} \over E_P + M_P}\end{array}\right) \chi_f
\label{spinor}
\end{equation}
where $\chi_{i,f}$ are two-component spinors. The spinor normalization
is such that $\overline{U}(\vec{q},s)U(\vec{q},s^\prime
)=2 M \delta_{ss^\prime}$.
We then obtain:
\begin{equation}
{\cal M}= G_F m_\pi^2 \sqrt{2M_\Lambda(E_P+M_P)}\left(
A\chi^\dagger_f\chi_i + B \chi^\dagger_f{\vec{\sigma}\cdot\vec{q} \over
E_P+M_P}\chi_i\right).
\label{glione}
\end{equation}
It is convenient to define:
\begin{eqnarray}
\hat{q} &=& {\vec{q} \over |\vec{q}|} \nonumber \\
s &=& A \nonumber \\
p &=& {|\vec{q}| \over E_P+M_P} B \nonumber \\
C &=& G_F m_\pi^2 \sqrt{2M_\Lambda(E_P+M_P)}.
\label{notation}
\end{eqnarray}
In terms of these quantities we can compute the decay distribution.
Squaring the matrix element:
\begin{equation}
|{\cal M}|^2 = C^2 \chi^\dagger_f(s+p \vec{\sigma}\cdot\hat{q})
\chi_i \chi_i^\dagger(s^* + p^* \vec{\sigma}\cdot\hat{q})\chi_f .
\label{matsq}
\end{equation}
We then make use of the projection operators:
\begin{equation}
\chi_{i,f}\chi^\dagger_{i,f} = \half (1+\vec{\sigma}\cdot\vec{\omega}_{i,f}),
\label{proop}
\end{equation}
and the identity $ \sigma_i\sigma_j = \delta_{ij}+i \epsilon_{ijk}\sigma_k$
to find after some algebra:
\begin{eqnarray}
|{\cal M}|^2 &=& {|C|^2 \over 2}\biggl( |s|^2 + |p|^2 \biggr)\biggl[1+
\gamma\vec{\omega}_i\cdot\vec{\omega}_f + (1-\gamma)\hat{q}\cdot\vec{\omega}_i
\hat{q}\cdot\vec{\omega}_f \nonumber \\
&+& \alpha \hat{q}\cdot(\vec{\omega}_i+\vec{\omega}_f)
+ \beta \hat{q}\cdot(\vec{\omega}_f\times\vec{\omega}_i)\biggr].
\label{msqsim}
\end{eqnarray}
We have introduced the notation:
\begin{eqnarray}
\alpha & \equiv& {2 {\rm Re} s^*p \over |s|^2 + |p|^2}  \nonumber \\
\beta & \equiv& {2 {\rm Im} s^*p \over |s|^2 + |p|^2} \nonumber \\
\gamma & \equiv& {|s|^2 - |p|^2 \over |s|^2 + |p|^2} .
\label{albega}
\end{eqnarray}
However, only two of these parameters are independent, since $\alpha^2
+\beta^2 + \gamma^2=1$. Sometimes  one finds in the literature the parameters
$\alpha$ and $\phi$, with $\beta=\sqrt{1-\alpha^2}\sin\phi$ and
$\gamma=\sqrt{1-\alpha^2}\cos\phi$. The non-leptonic hyperon decay
is therefore completely described by three observables: the total decay
rate, and two parameters determining the angular distribution. We take
the latter to be $\alpha$ and $\beta$.

The total decay rate is given by:
\begin{equation}
\Gamma = {|\vec{q}|(E_P+M_P) \over 4 \pi M_\Lambda}G_F^2 m_\pi^4
\left(|s|^2+|p|^2\right).
\label{drate}
\end{equation}
One way to interpret the parameter $\alpha$ follows from considering the
angular distribution in the case when the final baryon polarization
is not observed:
\begin{equation}
{d\Gamma \over d\Omega}={1 \over 16 \pi^2}{|\vec{q}|\over M_\Lambda}
G_F^2m_\pi^4(E_P+M_P)\left(|s|^2+|p|^2\right)
\left(1+\alpha \hat{q}\cdot \vec{\omega}_i\right).
\label{alonly}
\end{equation}
The polarization of the decay proton in the $\Lambda^0$ rest frame is given by:
\begin{equation}
\vec{\cal P}_P = {1 \over 1+\alpha\vec{\cal P}_\Lambda\cdot\hat{q}}
\biggl[\left(\alpha+\vec{\cal P}_\Lambda\cdot\hat{q}\right)\hat{q}
+\beta
\left(\vec{\cal P}_\Lambda\times\hat{q}\right)+\gamma\left(\hat{q}
\times\left(\vec{\cal P}_\Lambda\times\hat{q}\right)\right)\biggr].
\label{polp}
\end{equation}
{}From this expression we can relate $\beta$ to the proton polarization
in the direction perpendicular to the plane formed by
the $\Lambda$ polarization and
the proton momentum. Similarly, if the initial hyperon is unpolarized,
$\alpha$ gives us the polarization of the proton. From all these
expressions it is clear that $\alpha$ governs a $T$-even
correlation, whereas $\beta$ governs a $T$-odd correlation. According
to our general considerations, we will be able to construct a $\CP$-odd
observable using the parameter $\beta$ that does not vanish in the
absence of final state interactions. It will, however, be an observable
extremely difficult to measure, as it will require the measurement of the
polarization of both initial and final baryons. In practice, it is not
possible to measure the proton polarization, so this observable is not
useful for the reaction $\Lambda \ra p \pi^-$. However, if one studies
the decay chain:
\begin{equation}
\Xi^- \ra \Lambda^0 \pi^- \ra p \pi^- \pi^-
\label{chain}
\end{equation}
it is possible to study the correlation that gives $\beta_\Xi$, since
the polarization of the final baryon (in this case the $\Lambda$) can be
obtained by analyzing the angular distribution of the second decay.

\subsection{$\cpv$ and unitarity phases}

Even though we are discussing weak decays, the final state consists of
strongly interacting particles, so there will be strong rescattering
phases. These phases are responsible for the non-zero value of $\beta$
even in the absence of $\cp$ violation, in accordance with Eq.~\ref{tborn}.
It is convenient to analyze the
final pion-nucleon system in terms of isospin and parity eigenstates. In
that way we can have in mind the simple picture:
\begin{equation}
\Lambda^0 \stackrel{H_W}{\longrightarrow}\bigg(p\pi\bigg)^I_\ell
 \stackrel{H_S}{\longrightarrow}\bigg(p\pi\bigg)^I_\ell .
\label{naive}
\end{equation}
At $t=0$ the weak Hamiltonian induces the decay of the $\Lambda^0$
into a pion-nucleon system with isospin and parity given by $I,~\ell$.
This pion-nucleon system is then
an eigenstate of the strong interaction.
Furthermore, at an energy equal to the
$\Lambda$ mass, it is the only state with these quantum numbers. The
pion-nucleon system will then rescatter due to the strong interactions into
itself, and in the process pick up a phase $\delta^I_\ell$. This is
an example of what is known as Watson's theorem, and an excellent
discussion can be found in T.D.Lee's book \cite{tdlee}. We reproduce here
the main steps in the proof\cite{tdlee}.
The matrix elements that we need are, to
lowest order in the weak interactions:
\begin{equation}
A^I_\ell = \matel{(p\pi)^I_\ell}{H_W}{\Lambda^0}_m .
\label{weakamp}
\end{equation}
Both states are eigenstates of angular momentum with $J=\half$ and
$J_z = m$,
but rotational invariance implies that this matrix element is independent
of the quantum number $m$, so we will drop that subscript. The time evolution
of the final state due to the strong interactions is given by:
\begin{equation}
\ket{(p\pi)^I_\ell} = U^\dagger_S(\infty,0)\ket{(p\pi)^I_\ell,{\rm free}}.
\label{timev}
\end{equation}
Replacing Eq.~\ref{timev} into
Eq.~\ref{weakamp}, taking the complex conjugate, introducing
several factors of $1=U^\dagger_T U_T$, and finally using Eq.~\ref{wigner}:
\begin{eqnarray}
A^I_\ell &=& \matel{(p\pi)^I_\ell,{\rm free}}{U_S(\infty,0)H_W(0)}{\Lambda^0}
\nonumber \\
A^{I*}_\ell &=& \bra{(p\pi)^I_\ell,{\rm free}
}^*U_S^*(\infty,0) H_W^*(0)\ket{\Lambda^0}^*
\nonumber \\
&=&\bra{(p\pi)^I_\ell,{\rm free}
}^*U^\dagger_T\left(U_TU_S^*(\infty,0)U_T^\dagger\right)
\left(U_T H_W^*(0)U^\dagger_T\right)U_T\ket{\Lambda^0}^*
\nonumber \\
&=&\matel{(p\pi)^I_\ell,{\rm free}}{\T^{-1}\left(\T U_S(\infty,0)\T^{-1}\right)
\left(\T H_W(0)\T^{-1}\right)\T}{\Lambda^0}.
\label{subst}
\end{eqnarray}
But we know that angular momentum eigenstates transform according
to Eq.~\ref{timeang}, so:
\begin{eqnarray}
\T\ket{\Lambda^0}&=&(-)^{{1\over 2}+m}\ket{\Lambda^0}_{-m}\nonumber \\
\T\ket{(p \pi)^I_\ell,{\rm free}}_m &=& (-)^{{1\over 2}+m}\ket{(p\pi)^I_\ell,
{\rm free}}_{-m}.
\label{timestate}
\end{eqnarray}
Assuming time reversal invariance of both the strong and weak interactions we
thus find:
\begin{equation}
A^{I*}_\ell = \matel{(p\pi)^I_\ell,{\rm free}}{U_S(-\infty,0)H_W(0)}
{\Lambda^0}_{-m}.
\label{substn}
\end{equation}
Again, due to rotational invariance, we can drop the $-m$ subindex.We
can also use
\begin{equation}
U_S(-\infty,0)=U_S(-\infty,\infty)U_S(\infty,0)=S^\dagger U_S(\infty,0)
\label{transls}
\end{equation}
to finally obtain:
\begin{eqnarray}
A^{I*}_\ell &=& \matel{(p\pi)^I_\ell,{\rm free}}{S^\dagger U_S(\infty,0)
H_W(0)}{\Lambda^0} \nonumber \\
&=& \sum_n \matel{(p\pi)^I_\ell,{\rm free}}{S^\dagger}{n}\matel{n}
{U_S(\infty,0)H_W(0)}{\Lambda^0}.
\label{finalcc}
\end{eqnarray}
However, we have argued that at an energy equal to the $\Lambda^0$ mass,
the $p\pi$ state is the only state with quantum numbers $I,\ell$. The
$S$-matrix element, thus, vanishes for all elements of the sum except for
the case where the intermediate state is the same as the final state, in
which case it is just equal to the strong phase shift
\begin{equation}
\matel{(p\pi)^I_\ell,{\rm free}}{S^\dagger}{(p\pi)^I_\ell,{\rm free}}
=e^{-2i\delta^I_\ell},
\label{phasesh}
\end{equation}
and thus
\begin{equation}
A^{I*}_\ell = e^{-2i\delta^I_\ell}A^I_\ell .
\label{watson}
\end{equation}
This completes the proof of the statement that in the absence of $\CP$
violation, the phase of the decay amplitude is equal to the strong
rescattering phase of the final state. You already saw a similar result
in the analysis of the parameters $\epsilon$ and $\epsilon^\prime$ in
kaon decays.\cite{rafael}

Getting back to our problem, we can write the matrix elements for
the different isospin and parity channels as:
\begin{eqnarray}
s&=&s_1e^{i(\delta^1_s+\phi^1_s)}+s_3e^{i(\delta^3_s+\phi^3_s)}\nonumber \\
p&=&p_1e^{i(\delta^1_p+\phi^1_p)}+p_3e^{i(\delta^3_p+\phi^3_p)},
\label{isodef}
\end{eqnarray}
where the notation is such that: the indices $1,3$ refer to isospin $1/2,3/2$
final states; $\delta$ are the strong rescattering
phases; and $\phi$ are possible $\CP$ violating phases.
In this example, the final state with isospin $1/2$, $3/2$  can only be
reached via the $\Delta I= 1/2,3/2$ parts of the weak Hamiltonian.
According to the $\Delta I =1/2$ rule, therefore, the final state is
predominantly an isospin $1/2$ state.

We now relate this decay to the corresponding anti-particle decay
$\overline{\Lambda}^0\ra\overline{p} \pi^+$. To do this, we repeat the
analysis used to prove Watson's theorem, but without assuming time
reversal invariance of the weak interactions. We find:
\begin{eqnarray}
A^{I*}_\ell &=& e^{-2i\delta^I_\ell}\matel{(p\pi)^I_\ell}{
\T H_W\T^{-1}}{\Lambda} \nonumber \\
&=& e^{-2i\delta^I_\ell}\matel{(p\pi)^I_\ell}{(\cp)^{-1}(\CP\T)H_W
(\CP\T)^{-1}(\CP)}{\Lambda}
\label{cppred}
\end{eqnarray}
We now assume $\CP\T$ invariance of the weak interactions to find:
\begin{equation}
A^{I*}_\ell =e^{-2i\delta^I_\ell}\matel{(p\pi)^I_\ell}{(\cp)^{-1}H_W
(\CP)}{\Lambda}
\label{cptcc}
\end{equation}
The next step involves a choice of $\CP$ phases. It is easy to convince
yourself that the physics does not depend on this choice, although
intermediate steps might look different. We choose, in accordance
with Eq.~\ref{cpmeson}:
\begin{eqnarray}
\CP\ket{\Lambda}&=&\ket{\overline{\Lambda}} \nonumber \\
\CP\ket{(p\pi)^I_\ell}&=&-(-)^\ell\ket{(p\pi)^I_\ell}.
\label{phaseconv}
\end{eqnarray}
Defining the amplitudes for the anti-lambda decay via the notation
\begin{equation}
\overline{A}^I_\ell \equiv \matel{(\overline{p}\pi)^I_\ell}
{H_W}{\overline{\Lambda}}
\label{antidef}
\end{equation}
we then find\cite{marshak}:
\begin{equation}
A^{I*}_\ell = -(-)^\ell \overline{A}^I_\ell e^{-2i\delta^I_\ell} .
\label{rescpt}
\end{equation}
If, on the other hand, we assume $\CP$ invariance of the weak
Hamiltonian, we are led to
\begin{equation}
A^I_\ell = \matel{(p\pi)^I_\ell}{(\cp)^{-1}\left(\cp H_W \cp^{-1}\right)(\cp)}
{\Lambda}.
\label{cpinter}
\end{equation}
With the same phase convention as above, we thus find that $\cp$ invariance
would predict\cite{marshak}:
\begin{equation}
A^I_\ell = - (-)^\ell \overline{A}^I_\ell .
\label{cppredn}
\end{equation}
If we parameterize the amplitudes in non-leptonic hyperon decays as:
\begin{equation}
S =\sum_I s_I e^{i(\delta^I_s+\phi^I_s)}\;\;
P =\sum_I p_I e^{i(\delta^I_p+\phi^I_p)},
\label{newpar}
\end{equation}
then $\cp\T$ invariance of the weak Hamiltonian predicts:\cite{wolfen}
\begin{equation}
\overline{S} =\sum_I -s_I e^{i(\delta^I_s-\phi^I_s)}\;\;
\overline{P} =\sum_I p_I e^{i(\delta^I_p-\phi^I_p)},
\label{cpthyp}
\end{equation}
whereas $\cp$ invariance of the weak interactions predicts:
\begin{equation}
\overline{S} =\sum_I -s_I e^{i(\delta^I_s+\phi^I_s)}\;\;
\overline{P} =\sum_I p_I e^{i(\delta^I_p+\phi^I_p)}.
\label{cphyp}
\end{equation}
{}From this we see again that the $\phi^I_\ell$ phases violate $\cp$.
Our task now, is to compare the hyperon and anti-hyperon decay to
construct $\cp$ odd observables to extract $\phi^I_\ell$.

Once again we emphasize that
\begin{equation}
\beta = {2{\rm Im}s^*p \over |s|^2 + |p|^2} \not= 0
\label{betcag}
\end{equation}
even if $\cp$ is conserved. In our previous language, a (naive)-$T$-odd
observable is {\it not} a $\cp$ or $\T$ odd observable.

\subsection{Observables}

With the expressions of the previous section,
we can compute the three observables for the decay $\Lambda^0 \ra p \pi^-$:
\begin{eqnarray}
\Gamma &=& {|\vec{q}|C^2\over 8 \pi M_\Lambda^2}\biggl\{\sum_I\left(|s_I|^2 +
|p_I|^2\right) \nonumber \\
&+& \sum_{I\not= J}2s_Is_J\left[\cos\dfs\cos\dds-\sin\dfs\sin\dds
\right]\nonumber \\
&+&\sum_{I\not = J}2p_I p_J\left[\cos\dfp\cos\ddp-\sin\dfp\sin\ddp\right]
\biggr\}\nonumber \\
\alpha\Gamma &=&{|\vec{q}|C^2\over 4 \pi M_\Lambda^2}\sum_{I,J} s_Ip_J
\left[\cos\dfsp\cos\ddsp-\sin\dfsp\sin\ddsp\right]\nonumber \\
\beta\Gamma &=&{|\vec{q}|C^2\over 4 \pi M_\Lambda^2}\sum_{I,J}s_Ip_J
\left[\sin\dfsp\cos\ddsp+\cos\dfsp\sin\ddsp\right]
\label{obs}
\end{eqnarray}
whereas for the $\overline{\Lambda}^0\ra \overline{p} \pi^+$ decay we find:
\begin{eqnarray}
\overline{\Gamma}
&=& {|\vec{q}|C^2\over 8 \pi M_\Lambda^2}\biggl\{\sum_I\left(|s_I|^2 +
|p_I|^2\right) \nonumber \\
&+& \sum_{I\not= J}2s_Is_J\left[\cos\dfs\cos\dds+\sin\dfs\sin\dds
\right]\nonumber \\
&+&\sum_{I\not = J}2p_I p_J\left[\cos\dfp\cos\ddp+\sin\dfp\sin\ddp\right]
\biggr\}\nonumber \\
\overline{\alpha}\overline{\Gamma}
 &=&-{|\vec{q}|C^2\over 4 \pi M_\Lambda^2}\sum_{I,J} s_Ip_J
\left[\cos\dfsp\cos\ddsp+\sin\dfsp\sin\ddsp\right]\nonumber \\
\overline{\beta}\overline{\Gamma}
 &=&{|\vec{q}|C^2\over 4 \pi M_\Lambda^2}\sum_{I,J}s_Ip_J
\left[\sin\dfsp\cos\ddsp-\cos\dfsp\sin\ddsp\right]
\label{antiobs}
\end{eqnarray}
Comparing these expressions, we can construct the three $\cp$-odd
observables:\cite{donpa,he}
\begin{eqnarray}
\Delta &\equiv & {\Gamma - \overline\Gamma \over \Gamma + \overline\Gamma}
\nonumber \\
A & \equiv &{\alpha \Gamma + \overline{\alpha}\overline{\Gamma} \over
\alpha \Gamma - \overline{\alpha}\overline{\Gamma}}
= -{\sum_{I,J}s_Ip_J\sin\dfsp\sin\ddsp \over \sum_{I,J}s_Ip_J
\cos\dfsp\cos\ddsp}\nonumber \\
B &\equiv &{\beta\Gamma + \overline{\beta}\overline{\Gamma} \over
\beta\Gamma - \overline{\beta}\overline{\Gamma}}
=-{\sum_{I,J}s_Ip_J\sin\dfsp\cos\ddsp \over \sum_{I,J}s_Ip_J
\cos\dfsp\sin\ddsp}
\label{cpobs}
\end{eqnarray}
It is, perhaps, more useful to construct approximate expressions based
on the fact that there are three small parameters in the problem:
\begin{itemize}
\item The strong rescattering phases are measured to be small.
\item The $\Delta I=3/2$ amplitudes are much smaller than the
$\Delta I=1/2$ amplitudes.
\item The $\cp$ violating phases are small.
\end{itemize}
To leading order in all the small quantities one finds:\cite{he}
\begin{eqnarray}
\Delta &=& \sqrt{2}{s_3\over s_1}
\sin\left(\delta_s^3-\delta_s^1\right)
\sin\left(\phi_s^3-\phi_s^1\right)\nonumber \\
A&=& -\tan\left(\delta_p^1-\delta_s^1\right)
\sin\left(\phi_p^1-\phi_s^1\right)\nonumber \\
B&=& -\cot\left(\delta_p^1-\delta_s^1\right)
\sin\left(\phi_p^1-\phi_s^1\right)
\label{approxas}
\end{eqnarray}
We can see in these expressions that $\Delta$ arises mainly from an
interference between a $\Delta I=1/2$ and a $\Delta I=3/2$ $s$-waves,
and that it is suppressed by three small quantities. On the other
hand, $A$ arises as an interference of $s$ and $p$-waves of the same
isospin  and, therefore, it is not suppressed by
the $\Delta I =1/2$ rule. Finally, we can see that $B$ is not suppressed
by the small rescattering phases. This is as we expected for a $\cp$ odd
observable that is also (naive)-$T$ odd. The hierarchy $B >> A >> \Delta$
emerges.\cite{donpa}

This is as far as we can go in a model independent manner. If we want to
predict the value of these observables within a model for $\cp$ violation
we take the value of $s_3/s_1$ and the strong rescattering phases
from experiment and we try to compute the weak phases from theory.

\subsection{Standard model calculation}

In the case of the minimal standard model, the $\cp$ violating phase resides
in the CKM matrix. For low energy transitions, this phase shows up as the
imaginary part of the Wilson coefficients in the effective weak
Hamiltonian.\cite{wise}
In the notation of Buras \cite{buras},
\begin{eqnarray}
H_W^{eff} &=& {G_F \over \sqrt{2}}V^*_{ud}V_{us}\sum_i c_i(\mu)Q_i(\mu) +
{\rm ~hermitian~conjugate}\nonumber \\
Q_1 &=& (\overline{s}d)_{V-A}(\overline{u}u)_{V-A} \nonumber \\
Q_2 &=& (\overline{s}u)_{V-A}(\overline{u}d)_{V-A} \nonumber \\
Q_3 &=& (\overline{s}d)_{V-A}\sum_{q=u,d,s}(\overline{q}q)_{V-A} \nonumber \\
Q_5 &=& (\overline{s}d)_{V-A}\sum_{q=u,d,s}(\overline{q}q)_{V+A} \nonumber \\
Q_6 &=&-8 \sum_{q=u,d,s}(\overline{s}_Lq_R)(\overline{q}_Rd_L) \nonumber \\
Q_7 &=&{3\over 2} (\overline{s}d)_{V-A}
\sum_{q=u,d,s}e_q(\overline{q}q)_{V+A} \nonumber \\
Q_8 &=&-12 \sum_{q=u,d,s}e_q(\overline{s}_Lq_R)(\overline{q}_Rd_L)
\label{effweak}
\end{eqnarray}
The origin of the different terms in this Hamiltonian was discussed in the
lectures by De~Rafael. Here, I just remind you that the Wilson coefficients
are typically written as
\begin{eqnarray}
c_i(\mu) &=&z_i(\mu)+\tau y_i(\mu) \nonumber \\
\tau &=& - {V^*_{td}V_{ts} \over V^*_{ud}V_{us}}
\end{eqnarray}
and the $\cp$ violating phase is the phase of $\tau$.
Numerical values for these coefficients can be found, for
example, in Buchalla {\it et. al.}.\cite{buchalla}

The calculation would proceed as usual, by evaluating the hadronic matrix
elements of the four-quark operators in Eq.~\ref{effweak} to obtain real
and imaginary parts for the amplitudes, schematically:
\begin{equation}
\matel{p \pi}{H_w^{eff}}{\Lambda^0}|^I_\ell = {\rm Re}M^I_\ell + i {\rm Im}
M^I_\ell ,
\label{schematic}
\end{equation}
and to the extent that the $\cp$ violating phases are small, they can be
approximated by
\begin{equation}
\phi^I_\ell \approx { {\rm Im}M^I_\ell \over {\rm Re}M^I_\ell}.
\label{smallph}
\end{equation}
At present, however, we do not know how to compute the matrix elements so
we cannot actually implement this calculation. If we try to follow what is done
for kaon decays, we would compute the matrix elements using factorization
and vacuum saturation as a reference point, then define some parameters
analogous to $B_K$ that would measure the deviation of the matrix elements
from their vacuum saturation value. A reliable calculation of the ``$B$''
parameters would probably have to come from lattice QCD.

For a simple estimate, we can take the real part of the matrix
elements from experiment (assuming that the measured amplitudes are real,
that is, that $\cp$ violation is small), and compute the imaginary
parts in vacuum saturation. Since the vacuum saturation result
is much smaller than the measured amplitudes, this provides a conservative
estimate for the weak phases. There are many models in the literature that
claim to fit the experimentally measured amplitudes. Without entering into
the details of these models, it is obvious that to fit the data,
the models must enhance
some or all of the matrix elements with respect to vacuum saturation. Clearly,
one would get completely different phases depending on which matrix elements
are enhanced. It is not surprising, therefore, that a survey of these models
yields weak $\cpv$ phases that differ by an order of magnitude\cite{steger}.

To compute some numbers, we take from experiment:\cite{overseth,roper}
\begin{eqnarray}
{s_3 \over s_1} & = & -0.019 \pm 0.006    \nonumber \\
\delta^1_s = 6^\circ  && \delta^3_s = -3.8^\circ \nonumber \\
\delta^1_p = -1.1^\circ  && \delta^3_p = -0.7^\circ
\label{expph}
\end{eqnarray}
and the approximate weak phases estimated in vacuum saturation:\cite{steger}
\begin{eqnarray}
\phi^1_s &\approx& -3 y_6 {\rm Im}\tau \nonumber \\
\phi^1_p &\approx& -0.3 y_6 {\rm Im}\tau \nonumber \\
\phi^3_s &\approx& \left[3.56(y_1+y_2)+4.1(y_7+2y_8)
{m_\pi^2 \over m_s (m_u+m_d)}\right] {\rm Im}\tau
\label{approxph}
\end{eqnarray}
To get some numerical estimates we use the values for the
Wilson coefficients of Buchalla {\it et. al.}\cite{buchalla} with
$\mu=1$~GeV, $\Lambda_{QCD}=200$~MeV. Although quantities such as the
quark masses that appear in Eq.~\ref{approxph} are not physical\cite{dghb},
we will use for an estimate the value $m_\pi^2/(m_s(m_u+m_d))\sim 10$.
For the quantity ${\rm Im}\tau$ we use the upper bound from Eq.~\ref{jexp},
${\rm Im}\tau \leq 0.0014$. With all this we find:
\begin{eqnarray}
\Delta &=&\left\{ \begin{array}{ll}
-1.4 \times 10^{-6}& {\rm~for~} m_t=150~GeV \\
 -9.1 \times 10^{-7} &  {\rm~for~}m_t=200~GeV \end{array}\right.\nonumber \\
A &=& 3.7 \times 10^{-5} \nonumber \\
B &=& 2.4 \times 10^{-3}
\label{vsnumres}
\end{eqnarray}
A survey of several models for the hadronic matrix elements, combined
with a careful analysis of the allowed range for the short distance
parameters that enter the calculation yielded similar results: that
$A$ was in the range of ${\rm ``a~few"~}\times 10^{-5}$ and that
$\Delta$ was two orders of magnitude smaller. The rate asymmetry
exhibits a strong dependence on the top-quark mass: for a certain
value of $m_t$,
the two terms in Eq.~\ref{approxph} cancel
against each other. The angular
correlation asymmetries, on the other hand, depend mildly on the top-quark
mass. This is understood from the point of view that the most important
effect of a large top-quark mass is to enhance electroweak corrections
to the effective weak Hamiltonian. This is important for the $\Delta I=
3/2$ amplitudes but not for the $\Delta I =1/2$ amplitudes.

\section{$\cp$-violating effective Lagrangian}

In this section we show a few examples of $\cp$ violating new physics
that can be discussed in terms of a low energy effective interaction.
We use these interactions in the following section to estimate bounds that
different observables can place on the new physics. An extensive list of
$\cpv$ operators compatible with the symmetries of the standard model has
been given by Burgess and Robinson.\cite{burg} The importance of requiring
the effective operators to be gauge invariant has been emphasized by
de~R\'{u}jula and collaborators.\cite{rujula}

\subsection{Operators that can appear at tree-level}

\begin{figure}[htp]
\centering{\includegraphics[width=6cm]{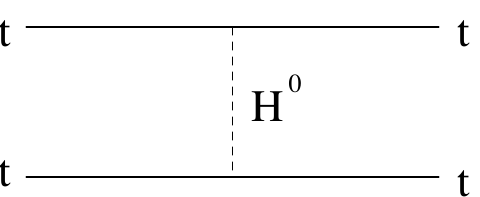}}
\caption{$\cp$ violating operators at tree-level.}
\label{treecp}
\end{figure}

An example of heavy particle exchange at tree-level, that
results in a $\cp$ violating effective Lagrangian is that
shown in Figure~\ref{treecp}.
Evaluating the diagram in Figure~\ref{treecp} with the couplings of
Eq.~\ref{neuyuk} we find:
\begin{equation}
{\cal M} = \left( -{m_t \over v}\right)^2\overline{t}
\left({\rm Re}A - i{\rm Im}A\gamma_5 \right)t \overline{t}
\left({\rm Re}A - i{\rm Im}A\gamma_5 \right)t {i \over q^2 - M^2_H}.
\label{fourt}
\end{equation}
In the heavy mass limit, this matrix element can be derived from the
effective interaction:
\begin{equation}
{\cal L}^{\cpv}_{eff} = i \sqrt{2}\left({m_t \over v}\right)^2
{1 \over M_H^2}\overline{t}\gamma_5 t \overline{t} t.
\label{lfourt}
\end{equation}
where we have assumed the unitarity upper bound for the $\cpv$ coupling.
Similarly, we can write down interactions like:
\begin{equation}
{\cal L}^{\cpv}_{eff} = i \overline{\eta}{g^2 \over \Lambda^2}
\overline{d}u\overline{t}\gamma_5 b + {\rm ~h.~c.}
\label{fourqc}
\end{equation}
obtained after Fierzing the operator that arises from an $s$-channel
exchange of a charged scalar of mass $\Lambda$ that couples with strength
$g$ and $\cpv$ mixing angles $\overline{\eta}$.

\subsection{One-loop effective operators: dipole moments}

One of the most commonly studied $\cpv$ operators that can appear at one-loop
is the electric-dipole moment of a fermion and its generalizations to weak and
strong couplings. The most general matrix element of the electromagnetic
current between two spinors contains a $\T$-odd term:
\begin{equation}
\matel{\Psi}{j^{em}_\mu}{\Psi}= i F_3(q^2)\overline{U}(p_2)\sigma_{\mu\nu}
q^\nu\gamma_5 U(p_1).
\label{toddjem}
\end{equation}
The value of this form factor at zero-momentum transfer:
\begin{equation}
F_3(q^2=0) \equiv d^\gamma_\Psi
\label{edmdef}
\end{equation}
is called the electric-dipole-moment. This induces a local interaction that
can be derived form the effective Lagrangian:
\begin{eqnarray}
{\cal L}^d_{eff} &=& {1 \over 2} d_f^\gamma \overline{\Psi}_f i
\sigma_{\mu\nu} \gamma_5 \Psi_f F^{\mu\nu} \nonumber \\
&+& {1 \over 2} d_f^Z \overline{\Psi}_f i
\sigma_{\mu\nu} \gamma_5 \Psi_f Z^{\mu\nu} \nonumber \\
&+& {1 \over 2} d_f^g \overline{\Psi}_f i
\sigma_{\mu\nu} \gamma_5{\lambda_a \over 2} \Psi_f G^{\mu\nu}_a
\label{edmleff}
\end{eqnarray}
where we have also added the generalizations to fermion couplings to the $Z$
boson and gluons.

Recalling that $\sigma_{\mu\nu}\gamma_5 = (i/2) \epsilon_{\mu\nu\rho\sigma}
\sigma^{\rho\sigma}$, it is easy to prove that this interaction is
indeed $\pp$ and $\T$ odd. Under parity:
\begin{eqnarray}
\pp A^\mu \pp^{-1} = A_\mu &&
\pp \partial^\mu \pp^{-1} = \partial_\mu \nonumber \\
\pp F^{\mu\nu} \pp^{-1} = F_{\mu\nu} && \pp
: \overline{\Psi}\sigma^{\mu\nu}\Psi:
\pp^{-1} = :\overline{\Psi}\sigma_{\mu\nu} \Psi:
\label{partrans}
\end{eqnarray}
and under time reversal invariance:
\begin{eqnarray}
\T A^\mu \T^{-1} = A_\mu &&
\T \partial^\mu \T^{-1} =- \partial_\mu \nonumber \\
\T F^{\mu\nu} \T^{-1} = -F_{\mu\nu} && \T : \overline{\Psi}\sigma^{\mu\nu}\Psi:
\T^{-1} =- :\overline{\Psi}\sigma_{\mu\nu} \Psi:
\label{timetrans}
\end{eqnarray}
using $\epsilon^{\mu\nu\rho\sigma}=-\epsilon_{\mu\nu\rho\sigma}$ it then
follows
that $d^\gamma_f$ is odd under both $\pp$ and $\T$.

\begin{figure}[htp]
\centering{\includegraphics[width=10cm]{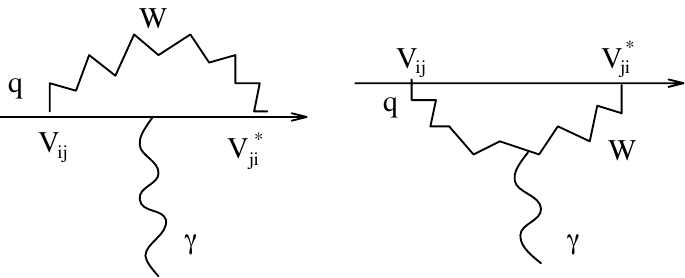}}
\caption[]{Potential contributions to a quark edm in the standard model at
one loop.}
\label{smoneedm}
\end{figure}

In the minimal standard model the electric dipole moment of a quark vanishes at
the one-loop order. Diagrams that might contribute are shown in
Figure~\ref{smoneedm}, where it is seen that at one-loop there can be
no $\cpv$ phase.\cite{chengli}
Diagrams at two-loop order can have a $\cpv$ phase, but it has been
shown by Shabalin \cite{shabalin} that the sum of two-loop
contributions to $F_3(q^2=0)$ vanishes. It is thus thought that in
the standard model, the lowest order contribution to the quark
electric dipole moment occurs at the three-loop level.

As an aside, it is worth commenting that the electric dipole moment
of the neutron is a much more complicated quantity to calculate
(as in many other cases the complication arises in computing the
hadronic matrix elements of quark operators). It seems, however,
that the value of the neutron EDM in the standard model is many
orders of magnitude below the current experimental upper bound.\cite{barr}
It has been claimed that it can be as large as $10^{-30}$~e-cm,\cite{gbarr}
although more likely values are at the $10^{-33}$~e-cm level.\cite{hbarr}
Recall that the experimental results for the neutron and electron
edm are:
\begin{eqnarray}
d_n^\gamma &=& (-1.4 \pm 0.6)\times 10^{-25}{\rm e-cm}
\nonumber \\
d_n^\gamma &=& (-0.3 \pm 0.5)\times 10^{-25}{\rm e-cm}
\nonumber \\
d_e^\gamma &=& (-2.7 \pm 8.3)\times 10^{-27}{\rm e-cm}
\label{expresedm}
\end{eqnarray}
where the numbers come respectively from experiments at
Leningrad,\cite{leningrad} Grenoble,\cite{grenoble} and
LBL.\cite{berkeley}.

\begin{figure}[htp]
\centering{\includegraphics[width=10cm]{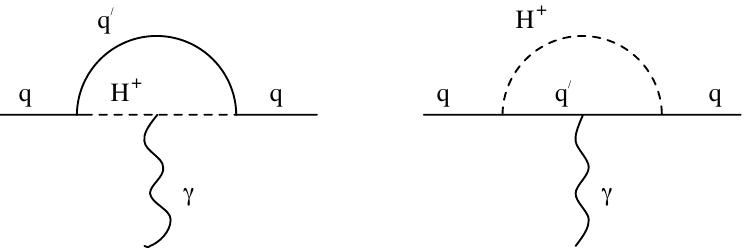}}
\caption{Quark edm from charged Higgs at one-loop.}
\label{chedm}
\end{figure}

There are models where it is possible to obtain
a non-zero quark electric-dipole
moment at the one-loop level. Examples are the models of $\cpv$ with extra
scalars. In the case where the $\cp$ violation arises in the charged scalar
sector, the electric-dipole moment is generated by the diagrams in
Figure~\ref{chedm}.
For down-type quarks, one obtains:\cite{edmchq}
\begin{equation}
d_D^\gamma = e{\sqrt{2}G_F \over 12 \pi^2}m_D {\rm Im}\alpha_1\beta_1^*
\left|V_{td}^*\right|{x_t \over (1-x_t)^2}\left({3 \over 4}-{5\over 4}x_t
+{1-{3\over 2}x_t \over 1-x_t}\log x_t\right)
\label{edmhc}
\end{equation}
where $x_t=m_t^2/M_H^2$. This result follows from the dominance of the
top-quark in the loop and assumes that the dominant contribution comes from
the lightest charged scalar $H^+$. For the case of an up-type quark the
result is:\cite{edmchq}
\begin{equation}
d_U^\gamma = e{\sqrt{2}G_F \over 12 \pi^2}m_U{\rm Im}\alpha_1\beta_1^*
\sum_i\left|V_{ui}\right|^2{x_i \over (1-x_i)^2}\left(x_i-
{1-3x_i \over 2(1-x_i)}\log x_i\right)
\label{edmhn}
\end{equation}
which is much smaller than the electric dipole moment of down-type quarks
due to the small masses of the quarks in the loop.

\begin{figure}[htp]
\centering{\includegraphics[width=6cm]{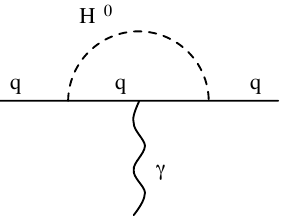}}
\caption{Fermion edm from neutral Higgs at one-loop.}
\label{nhedm}
\end{figure}

When $\cp$ violation comes from the exchange of a neutral Higgs, the quark
electric-dipole moment arises through the diagram of Figure~\ref{nhedm}.
In this case the value for up-type quarks,down-type quarks, or charged leptons
is given in the $M_H >> m_f$ limit by:\cite{edmnh}
\begin{equation}
d_f^\gamma = {eQ_f \sqrt{2}G_F \over 32 \pi^2}m_f^3{{\rm Im}A{\rm Re}A \over
M_H^2}\log\left({m_H^2\over M_f^2}\right).
\label{edmnhforq}
\end{equation}
This is largest for the top-quark (although in the case of the top-quark it
may be a poor approximation to take $q^2=0$).

\begin{figure}[htp]
\centering{\includegraphics[width=6cm]{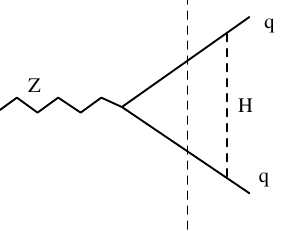}}
\caption{Possible absorptive phase in $d^Z_q$.}
\label{zedm}
\end{figure}

There are several new features that arise at high energy, for example, in the
generalization of the electric-dipole moments to couplings to the $Z$-boson.
For example, if $q^2 =0$ is not a good approximation, and the full form factor
is important, it is possible to have absorptive phases as shown schematically
in Figure~\ref{zedm}.
In this one-loop diagram, for $q^2 >> 4m_f^2$ the size of the imaginary
and real parts is comparable. In this type of problems there
is, therefore, no ``penalty'' associated with $\cp$-odd
observables that need unitarity phases, unlike previous
examples. In the limit $m_q << M_Z$, the diagram of Figure~\ref{zedm}
generates a $Z$-dipole moment:
\begin{eqnarray}
d_q^Z(q^2=M_Z^2)& =& -{g \over 4 \cos\theta_W}{g_V \over 16 \pi^2 v^2}
{\rm Re}A{\rm Im}A{m_q^3 \over M_Z^2}\nonumber \\
&&\cdot \left[{\rm Li}_2\left(-{q^2\over M_H^2}
\right) + \log\left(-{q^2 \over M_H^2}\right)\log\left(1+{q^2 \over M_H^2}
\right)\right]
\label{absedm}
\end{eqnarray}
Using, as always, the unitarity upper bound for the $\cpv$ phase, this
gives
\begin{equation}
{\rm Im}d^Z_b \approx {g \over 4 \cos\theta_W}
\left\{\begin{array}{ll}
10^{-8}~GeV^{-1} \sim 10^{-22}{\rm e-cm}&{\rm ~if~}M_H=M_Z\\
10^{-10}~GeV^{-1} \sim 10^{-24}{\rm e-cm}&{\rm ~if~}M_H = 10 M_Z\end{array}\right.
\label{absedmnum}
\end{equation}

As an aside, it is interesting to understand the origin of the
absorptive part of the dipole moments in the effective Lagrangian formalism,
where the couplings are really constants and not form
factors. The same interaction that generates the dipole moment at one-loop,
generates a $\cpv$ four fermion operator at tree-level as in Eq.~\ref{lfourt}.
In the effective
theory calculation at one-loop, one must include both the tree-level
$Z \ra \overline{f}f$ graph with the $d^Z_f$ vertex, and the one-loop
graph with a $Z\ra \overline{f}f$ standard model vertex followed
by a $\cpv$ four fermion vertex. It is this second diagram that contains
the absorptive part.
\newpage
\section{$\cp$-odd observables that probe $d^{\gamma,Z,g}_f$}

In this section we consider several observables that can be used to
place bounds on the couplings $d^{\gamma,Z,g}_f$.

\subsection{$\cp$-even observable}

The existence of a new coupling, even if it is a $\cp$-violating one,
changes the cross-section fermion pair production. By measuring
this cross-section precisely, one can place bounds on its deviations from
minimal standard model predictions and, thus, on new couplings like the
dipole moments. Barr and Marciano used this method to place
bounds on the electric dipole moment of the $\tau$.\cite{barr}
At PETRA energies, $\sqrt{s}=35$~GeV, the process $e^+ e^- \ra \tau^+ \tau^-$
is dominated by photon exchange. The cross-section from standard model and
electric-dipole moment couplings is given by:\cite{barr}
\begin{equation}
\sigma = {4 \pi \alpha^2 \over 3s}\left(1-{4m_\tau^2 \over s}\right)^{1\over 2}
\left(1+{2 m_\tau^2 \over s}\right)\left[1+{s \over 2}\left({d^\gamma_\tau
\over e}\right)^2\left(1-{4m_\tau^2 \over s}\right)\left(1+{2m_\tau^2
\over s}\right)^{-1}\right]
\label{sedm}
\end{equation}
Since no deviations from the standard model have been observed, Barr and
Marciano found that $|d^\gamma_\tau|\leq 1.3 \times 10^{-16}$~e-cm by
assuming that the cross-section can be measured to $5\%$. In order to
asses the significance of this bound, consider the case of
$\cp$ violation by exchange
of a neutral Higgs boson, Eq.~\ref{edmnhforq}.
Using the unitarity upper bound for the $\cpv$ phases, Eq.~\ref{unithn}, this
gives $d_\tau^\gamma \sim 10^{-23}$~e-cm.
The same formula, Eq.~\ref{edmnhforq},
applied to the electron gives $d^\gamma_e \sim 10^{-33}$~e-cm, which
is also much smaller than the current experimental bound, Eq.~\ref{expresedm}.

\subsection{$\cp$-even angular distribution}

It was shown by Del~Aguila and Sher~\cite{aguilasher}, that the differential
cross-section is also sensitive to the presence of an electric-dipole
moment of the tau. In this case, one does not have to assume a future
precision measurement of the cross-section, but can actually use what
has been measured by PETRA. The result is:\cite{aguilasher}
\begin{eqnarray}
{d\sigma \over d(\cos\theta)}&=&{\pi \alpha^2 \over 2s}\left[1+\cos^2\theta
+{1 \over 4 \sin^2\theta_W \cos^2\theta_W}{\rm Re}{s\over s-M_Z^2+i{\Gamma_Z
\over M_Z}}\right] \nonumber \\
&+&{1\over 2}\pi\alpha^2\left({d^\gamma_\tau \over e}\right)^2
\sin^2\theta ,
\label{dsdo}
\end{eqnarray}
Fitting the PETRA results for $\sqrt{s}=35$~GeV at the $1\sigma$~level,
the bound $|d^\gamma_\tau|\leq 1.4 \times 10^{-16}$~e-cm is
obtained.\cite{aguilasher}
\newpage
\subsection{$\cp$-odd angular correlations}

If we ignore the possibility of absorptive phases in the dipole-moment
form factors, and simply take $d^\gamma_\tau$ to be a constant, the
largest absorptive phases present in the reaction $e^+e^- \ra \tau^+
\tau^-$ come from the $Z$ width and are proportional to
$\Gamma_Z / M_Z << 1$. These are very small phases, and it therefore
becomes important to study $T$-odd observables, as was the case in
hyperon decays. Hoogeven and
Stodolsky\cite{hoog} studied the process
$e^+e^- \ra \tau^+ \tau^- \ra \pi^+ \overline{\nu}_\tau \pi^-
\nu_\tau$ and looked for the $\cpv$, $T$-odd correlation:
\begin{equation}
A= \left(\vec{p}_{e^+}-\vec{p}_{e^-}\right)\cdot\left(
\vec{p}_{\pi^+}\times \vec{p}_{\pi^-}\right).
\label{hoogcor}
\end{equation}
Since this correlation consists of a product of three momentum vectors
it is obviously $T$-odd. Recalling that in their center of mass, the initial
$e^+ e^-$ state transforms into itself under $\cp$ (Eq.~\ref{cpfpair}), and
using
\begin{equation}
\vec{p}_{\pi^\pm} \stackrel{\cp}{\longrightarrow} -\vec{p}_{\pi^\mp},
\label{cptrasmom}
\end{equation}
one sees that Eq.~\ref{hoogcor} is also $\cp$ violating.
This correlation is generated by
the interference between the amplitude proportional to the
electric-dipole moment of the tau and the
$Z$ exchange amplitude. It is, therefore, proportional to the real part of the
$Z$-exchange amplitude and vanishes on the $Z$ mass
shell. A numerical computation
of this asymmetry, scanning energies near the $Z$-mass
with an integrated luminosity equivalent to $10^7$ $Z$ bosons if running
on resonance, shows that one would be able to place a
$1\sigma$ limit $|d^\gamma_\tau|\leq 3 \times
10^{-16}$~e-cm.\cite{hoog}

\subsection{$\cp$-odd tensor observables}

Bernreuther and Nachtmann \cite{bern} showed that it is possible to
construct more complicated $\cp$-odd observables for the process
$e^+(\vec{p}_+)e^-(\vec{p}_-) \ra \tau^+(\vec{k}_+,\vec{s}_+)
\tau^-(\vec{k}_-,\vec{s}_-)$. Recalling from Eq.~\ref{cpfpair}, that under a
$\cp$ transformation this reaction goes into
itself with an interchange of the spin vectors
(in the $e^+ e^-$ center of mass frame ), it is easy to see
that the following two correlations are both $T$ and $\cp$ odd:
\begin{eqnarray}
A_{ij} &=& (\hat{k}_+)_i[\hat{k}_+\times(\vec{s}_- - \vec{s}_+)]_j
\nonumber \\
B_{ij} &=& (\hat{k}_+)_i[\vec{s}_-\times \vec{s}_+]_j
\label{tensorcor}
\end{eqnarray}
The first of these correlations is useful at the $Z$ resonance, whereas
the second one is useful at lower energies. In terms of the tensor
\begin{equation}
s_{ij} = {1 \over 2}(\hat{p}^+_i\hat{p}^+_j -{1\over 3}\delta_{ij})
\label{tensor}
\end{equation}
they found\cite{bern}:
\begin{eqnarray}
<A_{ij}> &=& {12 \sqrt{3} \over 5}\left({d^Z_\tau M_Z \over e}\right)
s_{ij} \nonumber \\
<B_{ij}> &=& -{12 \over 5}\left(1-{4m_\tau^2 \over s}\right)^{1\over 2}
\left(1+{4m_\tau \over 3 \sqrt{s}}\right)\left(1+{2m^2_\tau\over s}\right)^{-1}
\left({d^\gamma_\tau \over e}\right) \sqrt{s} s_{ij}
\end{eqnarray}
To construct more realistic observables one must specify how to measure the
tau polarization. One possibility is to study the tau decay
into a pion and a neutrino, where the angular distribution analyzes
the tau polarization.\cite{bern} Specifically, looking at the reaction
$e^+ e^- \ra \tau^+ \tau^- \ra \pi^+(q^+) \overline{\nu}_\tau \pi^-(q^-)
\nu_\tau$ one can construct the correlation:
\begin{equation}
T_{ij} = \left(\vec{q}_+ - \vec{q}_-\right)_i\left(\vec{q}_+ \times
\vec{q}_-\right)_j + \left ( i \tom j \right)
\label{eetauchain}
\end{equation}
which can easily be seen to be $T$ and $\cp$ odd. Numerically\cite{bern},
in units of $e$:
\begin{equation}
<T_{ij}> = \left\{\begin{array}{ll} -1.42 \times 10^{-3}({\rm GeV})^3 M_Z d^Z_\tau
s_{ij} &{\rm~for~}\sqrt{s}=M_Z\\
3.95 ({\rm GeV})^3 \sqrt{s} d^Z_\tau  s_{ij} &{\rm~for~}\sqrt{s}=10~GeV\end{array}\right. .
\label{bernnachres}
\end{equation}
With $10^7$ $Z$'s this could place the bound
$|d^Z_\tau|\leq 6 \times 10^{-18}$~e-cm. The same sensitivity can be
achieved with $10^6$ $\tau$ pairs in a $B$ factory at
$\sqrt{s}=10$~GeV.\cite{bern}

\subsection{$\cp$-odd, $\pp$ and $T$-even energy asymmetry}

\begin{figure}[htp]
\centering{\includegraphics[width=10cm]{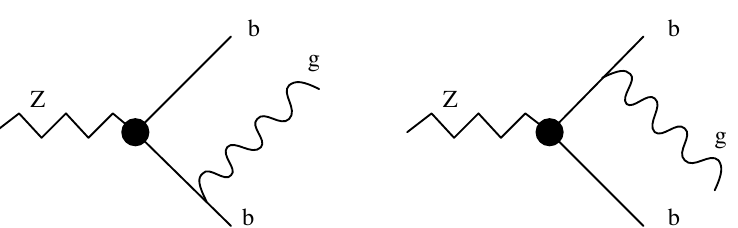}}
\caption{Three jet decay of the $Z$.}
\label{bedm}
\end{figure}

$\pp$-even observables in $Z$ decays are proportional to the vector
coupling of fermions to the $Z$ boson. Since this coupling
is accidentally small
for leptons, let us consider here the weak dipole moment of the
$b$-quark, $d^Z_b$. Specifically, we will consider a $T$-even
observable that requires final state interactions as an example of how these
can arise without additional suppression factors. We consider the
three jet decay of the $Z$ boson as depicted in Figure~\ref{bedm}.
The solid circle represents a vertex containing both the standard model
coupling and an absorptive weak dipole moment. If $\cp$ is conserved,
the average energy of the $b$ and $\overline{b}$ jets will be the same,
whereas they may differ if $\cp$ is violated. This is clearly a $T$-even
observable. We define it as:\cite{vaso}
\begin{equation}
{\cal O} = \left|{<E_b>-<E_{\overline{b}}>
\over <E_b>+<E_{\overline{b}}>} \right|
\label{obsvaso}
\end{equation}
Defining $x_b = 2E_b/M_Z$, $x_{\overline{b}}=2E_{\overline{b}}/M_Z$ and using
$g_V = 1 -4/3\sin^2\theta_W$, the tree-level partial width is:
\begin{equation}
\Gamma_0 \equiv \Gamma(Z\ra \overline{b}b)={\alpha M_Z \over 4 \sin^2
2\theta_W}
(1+g_V^2).
\label{treewidth}
\end{equation}
The three jet decay width in the standard model (at tree-level) is:
\begin{equation}
\Gamma(Z\ra \overline{b} b g)=\Gamma_0{2\alpha_S \over 3 \pi}\int
dx_b dx_{\overline{b}}{x_b^2 + x_{\overline{b}}^2 \over
(1-x_b)(1-x_{\overline{b}})},
\label{threejet}
\end{equation}
and the average $b$-jet energy is:
\begin{equation}
<E_b>={\Gamma_0 \over \Gamma(Z\ra \overline{b} b g)}{2\alpha_S \over 3\pi}
{m_Z \over 2}\int dx_b dx_{\overline{b}}{x_b(x_b^2 + x_{\overline{b}}^2) \over
(1-x_b)(1-x_{\overline{b}})}.
\label{aveenergy}
\end{equation}
We define the three-jet event by requiring that the invariant mass of any
pair of jets be larger than a minimum $5\%$ of the $Z$-mass. This value
corresponds to $<E_b> = 0.4 M_Z$. Using an absorptive ${\rm Im}d^Z_b$ as
computed before in Eq.~\ref{absedm}, we find for the interference between this
term and the standard model coupling:\cite{vaso}
\begin{equation}
{d\Gamma_{int} \over dx_b dx_{\overline{b}}}= M_Z {2 \alpha \alpha_s m_b
\over 3 \pi \sin^2 2 \theta_W}{(x_b - x_{\overline{b}})(x_b+x_{\overline{b}}
-2)\over (1-x_b)(1-x_{\overline{b}})}4\sin\theta_W\cos\theta_W{\rm Im}
{d^Z_b(M_Z^2) \over e}.
\label{vasoint}
\end{equation}
With this expression we find that the expectation value of the $\cp$-violating
correlation is:\cite{vaso}
\begin{equation}
<{\cal O}>\approx 0.1 m_b\left(4\sin\theta_W\cos\theta_W{\rm Im}
{d^Z_b(M_Z^2) \over e}\right)
\end{equation}
We can place a rough constraint by considering the case where the $b$ quark
becomes a $B$ or $B^*$ meson, and assuming that all the $b$-quark energy ends
up in the $B$-meson. Looking at $Z\ra B X$ events, there will be $6\times 10^5$
such events in a sample of $10^7$ $Z$ decays. With this number of events one
can place a $1\sigma$ limit $|{\rm Im}d^Z_b(M_Z^2)|
\leq 10^{-17}$~e-cm.\cite{vaso}
Once again, recall that with the unitarity upper bound for $\cp$ violating
phases, the exchange of a neutral Higgs generates this coupling at the
$|{\rm Im}d^Z_b(M_Z^2)|\sim 10^{-22}$~e-cm level.

\subsection{$\overline{t}t$ production in colliders}

\begin{figure}[htp]
\centering{\includegraphics[width=10cm]{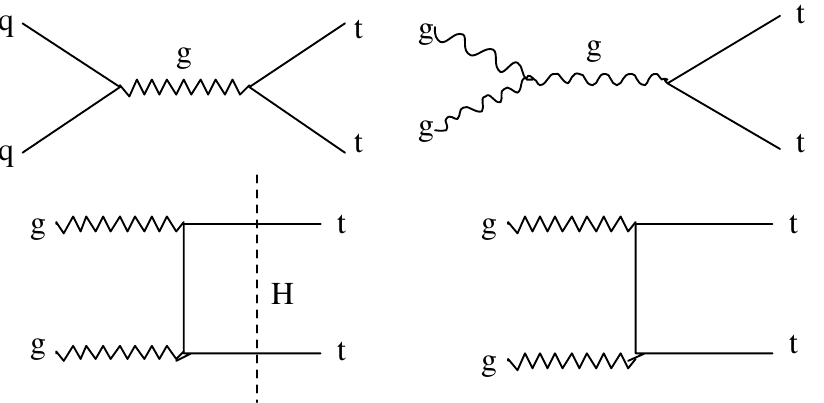}}
\caption[]{Some diagrams contributing to $\overline{t}t$ production in hadron
colliders}
\label{fpes}
\end{figure}

We can pursue the line of argument that has been followed throughout the
previous examples and ask for the largest possible signal of this type.
Clearly, we want to look at the heaviest fermion, the top-quark, since the
dipole-moment form factors are proportional to the third power of the
fermion mass. Also, we would get a larger form factor if we look at the
color-electric-dipole moment where the coupling is the strong coupling
constant instead of $e$. For the high energies needed to look at the
effect of a dipole-moment interaction in the production of a fermion pair,
the $q^2=0$ limit is not a good approximation so we can't really use the
language of dipole-moments anymore. Nevertheless the origin of the
$\cp$ violating interaction is the same. Peskin and Schmidt\cite{peskin}
have considered the production of top-quark pairs near threshold in
hadron colliders. This would happen through diagrams as those
in Figure~\ref{fpes}.
Helicity conservation implies that at very high energies, $E >> m_t$, one
will mostly produce $\overline{t}_Lt_R$ and $\overline{t}_Rt_L$ pairs.
However, near threshold, there is significant production of
$\overline{t}_L t_L$ and $\overline{t}_R t_R$ pairs. Under a $\cp$
transformation
\begin{equation}
t(\vec{p}_t,\vec{s}_t) \overline{t}(\vec{p}_{\overline{t}}
,\vec{s}_{\overline{t}}) \stackrel{\cp}{\longrightarrow}
 \overline{t}(-\vec{p}_t,\vec{s}_t) t(-\vec{p}_{\overline{t}},
\vec{s}_{\overline{t}})
\label{cpforpes}
\end{equation}
so
\begin{equation}
\overline{t}_L t_L \stackrel{\cp}{\longleftrightarrow} \overline{t}_R t_R
\label{cpconjpes}
\end{equation}
Thus, we can consider the $\cp$ violating observable:\cite{peskin}
\begin{equation}
\Delta N_{LR} = {N(\overline{t}_L t_L)-N(\overline{t}_R t_R) \over
N(\overline{t}_L t_L)+N(\overline{t}_R t_R)}
\end{equation}
This is a $T$ even observable that needs absorptive phases, however, as in the
previous example, they may be generated at the one-loop level with
no additional suppression factors as shown schematically in Figure~\ref{fpes}.
The resulting $\cp$ violating amplitude is closely related to the form factor
${\rm Im}d^g_t(s)$. As before, we must specify a way to look at the spin
information of the top-quark. In this case, the top-quark is sufficiently heavy
that it will decay weakly and its decay into $b W$ will analyze the
polarization.\cite{bigi}
Peskin and Schmidt argue that the lepton energy in the
subsequent decay $W \ra \ell \nu$ retains  information on the polarization
of the parent top-quark. Assuming the unitarity upper bound for the
$\cp$ violating phases, they find numerically that $|\Delta N_{LR}| \sim
10^{-3}$ is possible, and that at the level of lepton energy,
it is still possible to find asymmetries of order $10^{-3}$.\cite{peskin}

\section{Inclusive tests of $\cp$ violation}

Finally, we mention some ``inclusive'' tests of $\cp$ violation that
have been described in the literature.\cite{donva,kamion}
The advantage of tests like these,
is that they do not require complete flavor identification, crucial to most
of the tests we have described so far, and that is very difficult to achieve in
high energy experiments. One example considered in Ref.~\cite{donva} involves a
comparison of the processes:
\begin{equation}
e^+(p^+)e^-(p^-) \ra t \overline{t} g \left\{
\begin{array}{ll}\ra b \ell \nu
\overline{t} g & \\
\ra t \overline{b}\overline{\ell}\overline{\nu} g & \end{array}\right.
\label{donvapro}
\end{equation}
Clearly, if $\cp$ is conserved, these two processes have the same rate. In the
$e^+ e^-$ center of mass frame, a $\cp$ transformation on the first
reaction gives:
\begin{eqnarray}
&&e^+(\vec{p}^+)e^-(\vec{p}^-) \ra  b(\vec{p}_b) \ell(\vec{p}_\ell) \nu
\overline{t}(\vec{p}_t) g \stackrel{\cp}{\longrightarrow} \nonumber \\
&&e^+(\vec{p}^+)e^-(\vec{p}^-) \ra  t(-\vec{p}_t)
\overline{b}(-\vec{p}_b) \overline{\ell}(-\vec{p}_\ell) \overline{\nu} g
\label{whathap}
\end{eqnarray}
We can construct the two observables
\begin{eqnarray}
J_1&=& \vec{p}_{\overline{t}}\cdot\vec{p}_b\times \vec{p}_\ell \nonumber \\
J_2 &=& \vec{p}_t\cdot\vec{p}_{\overline{b}}\times \vec{p}_{\overline{\ell}}
\label{donvaobser}
\end{eqnarray}
and note that under $\cp$
\begin{equation}
J_1 \stackrel{\cp}{\longleftrightarrow} -J_2
\label{onetotwo}
\end{equation}
If the differential cross-section for the first reaction has a correlation
of the form $(d\sigma_1 /d\Omega) \sim  a + b J_1$, the second reaction will
have a correlation of the form $(d\sigma_2 /d\Omega) \sim  a + b J_2$. If we
can not tell apart the two reactions, but we know that they occur with equal
probability, $\cp$ invariance requires that there be no correlation of the
form $ J= \vec{p}_t\cdot\vec{p}_b\times\vec{p}_\ell$ in the inclusive
process.

This type of idea can be exploited to construct $\cp$-odd observables
that require no flavor identification. For example, one can look at
the reaction $e^+e^- \ra 4 {\rm ~jets}$, and order the four jets in any
$\cp$-blind way. For example, by ``fastness'' or ``fatness''. The $T$-odd
observable
\begin{equation}
{\cal O} = \vec{p}_1\cdot \vec{p}_2 \times\vec{p}_3
\label{jeto}
\end{equation}
is, thus, also a $\cp$-odd observable.
Using a simple model, with an interaction
of the form Eq.~\ref{fourqc}, it has been found that asymmetries of order
$10^{-3}$ are possible.\cite{donva}
Clearly, searching for these asymmetries is a
worthwhile enterprise, even though it would be difficult to interpret
an observation of $\cp$ violation through a non-zero expectation value of an
observable  such as Eq.~\ref{jeto} in terms of specific models of $\cp$
violation.

\section{Conclusions}

$\cp$ violation remains one of the unexplained aspects of particle
physics, and has been observed only in the decays of $K_L$. In order
to elucidate the origin of $\cp$ violation, it is crucial to observe
it in other systems. In systems more complicated than the $K^0 \ra
\pi \pi$ reaction, it is possible to construct many $\cp$-odd observables.
They have different characteristics and probe different aspects of
potential $\cp$ violating physics. In these lectures we have reviewed
the basic ingredients that go into the construction of $\cp$-odd
observables and we have sampled some of the proposals in the
literature. Clearly, the subject of searching for $\cp$ violation is
a vast one, and we cannot discuss it all in these lectures. The
selection of topics presented here was, of course, biased by my own
work in the field. Additional topics discussed recently that we did not
have time to mention include $\cpv$ at high energy colliders;\cite{high}
using $\gamma \gamma$ colliders with polarized photons;\cite{photons}
studying the production of $W^+W^-$ pairs;\cite{goun} and additional
observables in $e^+e^-$ collisions.\cite{other}

\section{Acknowledgements}

I am grateful to J. F. Donoghue for many discussions
on the subject of $\cp$ violation. I also wish to thank the organizers
of the school, J.~F.~Donoghue and K.~T.~Mahanthappa for their hospitality.


\begin{thebibliography}{999}
\bibitem{wolfen}{A similar discussion for observables in $B$ decays
can be found in L.~Wolfenstein, \prd{43}{151}{91}.}
\end{thebibliography}
\end{document}